\documentclass[11pt]{article}
\usepackage[margin=1in]{geometry}
\usepackage{yck}
\usepackage{tabularx}
\usetikzlibrary{arrows.meta,positioning,calc,fit,shapes.geometric}
\setcitestyle{authoryear,round}

\graphicspath{{fig/}}

\def\bc{\bm c}

\def\bw{\mathbf w}

\allowdisplaybreaks

\title{Single and multi-objective optimal designs for group testing experiments with a focus on screening for an infectious disease}

\author[1,2,3]{Chi-Kuang Yeh~\yckorcid}
\author[4]{Weng Kee Wong~\orcidlink{0000-0001-5568-3054}\thanks{Corresponding author. Email: \texttt{\href{mailto:wkwong@ucla.edu}{wkwong@ucla.edu}}.}}
\author[5]{Julie Zhou}

\affil[1]{Department of Mathematics and Statistics, Georgia State University}
\affil[2]{Department of Mathematics and Statistics, McGill University}
\affil[3]{Department of Statistics and Actuarial Science, University of Waterloo}
\affil[4]{Department of Biostatistics, University of California, Los Angeles}
\affil[5]{Department of Mathematics and Statistics, University of Victoria}

\date{\today}

\begin{document}

\maketitle

\begin{abstract}
Group testing techniques are widely used in resource-constrained settings, such as infectious-disease screening, blood safety, DNA library screening, and industrial inspection, where the efficient use of limited testing resources depends critically on how the initial study is designed. This paper discusses various ways that group testing experiments can be designed more efficiently and flexibly, under a user-specified optimality criterion and cost structure. We construct optimal designs to estimate model parameters beyond the \(D\)-optimality criterion to include the \(A\)-, \(c\)-, \(E\)-optimality, and extend the framework for finding optimal designs with multiple objectives. For large studies, we use a general theory and obtain various types of optimal approximate designs.  When sample sizes are small, we propose two algorithms to construct highly efficient exact designs under realistic budget constraints. Additionally, we investigate properties of the proposed designs under various operational uncertainties and create a Shiny app to facilitate implementation of the proposed designs. To fix ideas, we focus on finding highly efficient group testing designs for a Chlamydia screening trial with imperfect assays under budget constraints and show the advantages of our optimal designs over current methods. 
\end{abstract}


% \noindent\textbf{AMS subject classifications:} 62K05; 62J12; 62P10

\section{Introduction}\label{sec-introduction}

\textcolor{black}{Group testing, or pooled-sample testing, has a long history and remains an active area of research.  Since the seminal work of \citet{gtfirst}, there has been continuing statistical methodology development and they include  \cite{hwang1,hwang2,mcmahan:2012:informative, huang:2017:design-group,huang:2020:cost-GT,huangejs}. It is especially useful when individual testing is costly and the disease or defect of interest is rare \citep{hughes:1994:group,hughes:2000:group}.  Group testing methodology is increasingly used in different types of applications and some recent ones are in biosecurity \citep{biosec}, computer science \citep{compsci2,compsci1}, cybersecurity \citep{cyber}, epidemiology \citep{epid}, cryptography \citep{compsci4,compsci3}, engineering \citep{community}, public health \citep{covid1,covid2}. in DNA library screening \citep{ngo:2000:survey},\citet{haber:2021:prime-time} review the breadth of these applications and emphasize the continuing practical relevance of group testing across scientific and societal settings. Earlier work has also studied in the design of group testing experiments and related practical settings \citep{mitchell:1987:computer,balzer:2025:multistage}, and in multi-access communication protocols \citep{wolf:1985:principles}. \iffalse Recent developments further underscore the practical relevance  of group testing. For example, pooled PCR and matrix-based screening strategies have been developed for large-scale pathogen detection \citep{saeedi:2022:PCR-screening}, and
The main statistical challenges are to determine the optimal number of groups and how to allocate the limited testing resources efficiently across the groups to attain maximum inferential accuracy, often in the presence of imperfect tests, operational constraints, and competing study objectives.\fi}
\iffalse
\wpink{Recent developments further underscore the practical relevance \textcolor{black}{of group testing}. For example, pooled PCR and matrix-based screening strategies have been developed for large-scale pathogen detection \citep{saeedi:2022:PCR-screening}, and modern overviews continue to highlight the importance of group testing for prevalence estimation and efficient screening under limited resources \citep{bilder:2019:group}. Yet these developments do not remove the need for principled design of the initial study. On the contrary, when prevalence and assay characteristics must be estimated under budgetary and logistical constraints, the quality of the downstream testing procedure depends directly on how well this first-stage design problem is handled. This is precisely the setting in which optimal design methods can provide practical value.}\fi

In many modern applications, however, group testings  \textcolor{black}{are invariably} conducted under imperfect assays, heterogeneous costs, and other operational constraints. Much of the literature in   group testing has focused on estimation, classification, and downstream testing procedures once data have been collected \citep{model1,model2,model3}. Substantially less attention has been paid to the upstream design problem of how group sizes and resources should be chosen before data collection, especially when the study is subject to realistic cost constraints and multiple inferential objectives. This gap is important because the performance of downstream group testing procedures depends directly on how well key parameters, such as prevalence, sensitivity, and specificity, are estimated at the design stage.

From a design perspective, the central questions are how to choose group sizes and how to allocate resources across them under a specified statistical model, optimality criterion, and cost structure. These questions are distinct from the downstream analysis problems more commonly studied in the group testing literature, yet they are fundamental when the aim is to obtain reliable parameter estimates under limited resources. Exact optimal designs are often difficult to derive analytically because of their discrete and combinatorial nature. When the sample size is sufficiently large, a natural alternative is to consider \emph{optimal approximate designs} (OADs), which allocate proportions of experimental effort over the design space. OADs are attractive because they are both tractable and amenable to rigorous verification through equivalence theorems. For example, \citet{huang:2017:design-group} studied OADs for group testing by jointly determining group sizes and allocation proportions under several design criteria. \citet{li:2017-correlated-GT} also considered design issues for parameter estimation, but in a different modeling framework and under a more restricted class of designs.

The perspective taken in this paper is that precise parameter estimation should be viewed as a foundational first stage of group testing, rather than as a secondary step. Many practical group testing procedures are ultimately aimed at reducing the total number of assays, but this objective is typically pursued only after the underlying prevalence and assay characteristics have been estimated with sufficient precision. Our focus is therefore not on minimizing the number of tests directly, but on constructing designs that estimate the governing model parameters efficiently and robustly. As illustrated in Figure~\ref{tikz:sequential}, we regard this as \textbf{Stage I} of a two-stage framework, with subsequent sequential group testing procedures forming \textbf{Stage II}. A substantial amount of work has focused on what we refer to here as Stage II, including \citet{mcmahan:2012:informative}, \citet{model3}, and related contributions in the group testing literature.

\begin{figure}[ht!]
\centering
\begin{tikzpicture}[
line width=0.9pt,
font=\sffamily,
>=Stealth,
node distance=8mm
]

% ==== COLORS ====
\definecolor{stageblue}{RGB}{142,180,227}
\definecolor{stageorange}{RGB}{252,176,64}
\definecolor{stagegreen}{RGB}{120,190,120}
\definecolor{stagepurple}{RGB}{176,160,220}
\definecolor{stagegray}{RGB}{230,230,230}

\tikzset{
    longarrow/.style={-{Latex[length=3mm,width=2mm]}, line width=0.9pt},
  module/.style={
    draw,
    rounded corners=4pt,
    thick,
    minimum width=24mm,
    minimum height=10mm,
    inner sep=2pt,
    align=center,
    fill=white
  },
  smallbox/.style={
    draw,
    rounded corners=3pt,
    thick,
    minimum width=30mm,
    minimum height=9mm,
    inner sep=2pt,
    align=center,
    fill=white
  },
  rbox/.style={
    draw,
    rounded corners=3pt,
    thick,
    inner sep=2pt,
    minimum width=17mm,
    minimum height=8mm,
    align=center,
    fill=white
  },
  leaf/.style={
    draw,
    rounded corners=2pt,
    thick,
    inner sep=1pt,
    minimum width=13mm,
    minimum height=6mm,
    align=center,
    fill=white
  }
}

% ======================================================================
%  STAGE I PANEL
% ======================================================================
\draw[dashed, rounded corners=10pt, stageblue, very thick]
  (0,-1.0) rectangle (7.6,7.6);
\node[anchor=west, font=\bfseries] at (0.35,7.2)
  {Stage I: Parameter Estimation};

% \node[smallbox, fill=stagegray] (model) at (1.8,5.9)
%   {Pre-specified\\ model};

% \node[smallbox, fill=stagegray] (nominal) at (5.0,5.9)
%   {Nominal value\\ $\btheta_0$};

  \node[smallbox, fill=stagegray,
      text width=1.9cm,
      minimum width=2.1cm,
      minimum height=0.75cm,
      inner sep=2pt] (model) at (1.8,5.9)
  {Pre-specified\\ model};

\node[smallbox, fill=stagegray,
      text width=1.9cm,
      minimum width=2.1cm,
      minimum height=0.75cm,
      inner sep=2pt] (nominal) at (5.0,5.9)
  {Nominal value\\ $\btheta_0$};

\node[module, fill=stageorange!40] (design) at (1.8,3.5)
  {Locally optimal\\ design $\xi^*(\btheta_0)$};

\node[module, fill=stagegreen!35] (data) at (5.0,3.5)
  {Data};

\node[smallbox, fill=stagepurple!18] (update) at (3.4,1.2)
  {MLE \\ estimate $\hat{\btheta}$};

% arrows
\draw[->, thick] (model.east) -- (nominal.west);
\draw[->, thick] (nominal.south) -- (design.north);
\draw[->, thick] (design.east) -- (data.west);
\draw[->, thick] (data.south) -- (update.north);

% feedback loop: updated estimate used as next nominal value
\draw[->, thick] (update.north) --(design.south);

\node[font=\small] at (1.5,2.15) {Use $\hat{\btheta}$ as new $\btheta_0$};

\node[font=\small\itshape, align=center] at (3.8,-0.55)
  {Iterative locally optimal design--data \\update loop for estimating $\btheta$};

% Arrow to Stage II
\draw[longarrow]
  (7.6,3.9) -- node[above, align=center]{Stable\\[-1mm]$\boldsymbol{\theta}_0$} (8.6,3.9);
% \draw[longarrow]
  % (7.6,3.9) -- node[above]{Stable \\ $\hat{\btheta}_0$} (8.6,3.9);

% ======================================================================
%  STAGE II PANEL
% ======================================================================
\draw[dashed, rounded corners=10pt, stageblue, very thick]
  (8.6,-1.0) rectangle (15.4,7.6);
\node[anchor=west, font=\bfseries] at (8.9,7.2)
  {Stage II: Sequential Group Testing};

\node[rbox, fill=stageorange!40] (pool) at (12.0,5.4) {Subjects};

\node[rbox, fill=stageblue!20] (g1) at (10.0,4.0) {Pool 1};
\node[rbox, fill=stageblue!20] (g2) at (12.0,4.0) {Pool 2};
\node[rbox, fill=stageblue!20] (g3) at (14.0,4.0) {Pool 3};

\draw[->, thick] (pool.south) -- ++(0,-0.35) -| (g1.north);
\draw[->, thick] (pool.south) -- ++(0,-0.35) -- (g2.north);
\draw[->, thick] (pool.south) -- ++(0,-0.35) -| (g3.north);

\node[leaf, fill=stagegreen!25] (ind1) at (10.0,2.25) {Individual\\ tests};
\node[leaf, fill=stagegreen!25] (ind2) at (12.0,2.25) {Individual\\ tests};
\node[leaf, fill=stagegreen!25] (ind3) at (14.0,2.25) {Individual\\ tests};

\draw[->, thick] (g1.south) -- ++(0,-0.3) -- (ind1.north);
\draw[->, thick] (g2.south) -- ++(0,-0.3) -- (ind2.north);
\draw[->, thick] (g3.south) -- ++(0,-0.3) -- (ind3.north);

\node[module, fill=stagepurple!25, minimum width=30mm] (decision)
  at (12.0,0.9) {Decisions \& outcomes};

\draw[->, thick] (ind2.south) -- (decision.north);

\node[font=\small\itshape, anchor=west] at (8.9,-0.55)
  {Use $\hat{\btheta}_{\mathrm{stable}}$ for group testing to reduce assays};

\end{tikzpicture}
\caption{Overview of the two-stage framework. \textbf{Left}: Stage I represents an iterative locally optimal design procedure. A model is first specified with an unknown parameter vector \(\btheta\). Prior information or pilot data provide a nominal value \(\btheta_0\), which is used to construct a locally optimal design \(\xi^*(\btheta_0)\). Data collected under this design are then used to compute the maximum likelihood estimate \(\hat{\btheta}\), which serves as the updated nominal value \(\btheta_0\) for the next iteration. This process is repeated until the estimate stabilizes. \textbf{Right}: The resulting stable estimate \(\btheta_0\) is then used in Stage II to carry out sequential group testing with the aim of reducing the total number of assays. Subjects are partitioned into pooled groups, followed when necessary by individual tests and final diagnostic or classification outcomes.}
\label{tikz:sequential}
\end{figure}

 In this paper, there are four main contributions for finding optimal or highly efficient group testing designs under realistic fixed-sample-size and budget constraints. First, we broaden the optimality criteria for group testing designs to include  \(A\)-, \(c\)- and \(E\)-optimality, beyond the commonly studied and arguably overused \(D\)- and \(D_s\)-criteria. Second, we develop a unified theoretical framework for finding and verifying single and multiple-objective design problems, motivated by the practical need to balance competing inferential priorities in screening studies.  Third, we also consider the case where the sample sizes are small and the theory may no longer apply.  In this case, we propose two modified rounding algorithms that guarantee the group testing exact design is highly efficient.  Fourth, we create an R package, \texttt{gtDesign} \citep{gtDesign}, for computing optimal designs for group testing experiments  available at \url{https://github.com/chikuang/gtDesign} and on CRAN. 
%\textcolor{blue}{(I think readers are less persuaded by the usefulness of using $A, D, A_s$ or $D_s$ criteria because we did not provide advantages of using A over D-optimality in group testing; so may be helpful we tried to relate them to model mis-specification issues. Can we elaborate on this point if possible?)} 
 
These contributions add values to the current methodology for constructing group testing designs in the following ways.  The first contribution reflects different inferential priorities that can arise in practice. While \(D\)-optimality targets overall precision through the generalized variance, \(A\)-optimality is more directly related to average marginal precision, and \(E\)-optimality protects against having  poorly estimated estimators. Similarly, \(c\)-optimality is appropriate when scientific or operational interest centres on a specific linear combination of prevalence, sensitivity, and specificity. In group testing studies, such distinctions matter because different inferential priorities can lead to materially different designs and, as shown later, can also affect downstream testing decisions.  As best as we know, \(A\)- and \(E\)-optimality criteria are new for group-testing problems and so provide additional choices for the practitioners.  The second contribution constructs an optimal design for group testing experiments when there the multiple objectives. This seems to be an under-addressed problem in group testing trials. A distinguishing feature of our work is that we have a rigorous theoretical framework for finding and verifying optimality of a design for group testing problems with multiple objectives. An  example of a group testing design that has two objectives was given in \citet{sars2}, where the authors wanted to analyze the group testing strategy that maximizes the efficiency of the SARS-CoV-2 screening test while ensuring its effectiveness, where the effectiveness of group testing guarantees that negative results from pooled samples can be considered presumptive negative. The above theoretical approach uses asymptotic results and may not apply when we have small samples, and even if does, the design may not be implementable.  To this end, we propose two additional ways for finding efficient and implementable group testing designs for small sample studies.  To facilitate use of our methodology, our fourth contribution is a repository that contains additional numerical results that substantiate our findings, along with computer codes that can generate all results in our paper, and additionally, they can be amended to find other types of group testing designs under similar models.

Our proposed methodology is quite general and applies broadly across different types of group testing problems. To fix ideas, we illustrate the proposed method for constructing an optimal group testing design for screening  Chlamydia , a  sexually transmitted infection, using imperfect diagnostic assays under controlled budget constraints. The application shows that different design criteria can lead to markedly different estimation and resource-allocation properties, and  highlights the practical importance of Stage I design before implementing sequential group testing strategies.

Section~\ref{sec:application} introduces the Chlamydia screening problem and the real-world considerations that motivate the design formulation. Section~\ref{sec:OD} develops the methodology for optimal group testing designs. Section~\ref{sec:results} presents selected optimal group-testing designs for the Chlamydia application for large and small sample sizes.  We also investigate consequences of mis-specifications in the values of the parameters in the early stage affects statistical efficiency in Stage II. Section~\ref{sec:discussion} discusses extensions of the framework and investigates robustness properties of the optimal designs under alternative operational constraints.   Section~\ref{sec:conclusion} concludes the paper with a brief summary.

\iffalse An R package, \texttt{gtDesign}, for computing optimal designs for group testing experiments is available at \url{https://github.com/chikuang/gtDesign}. The repository also contains additional numerical results and fully reproducible code for all analyses reported in this paper.\fi

\section{Testing Chlamydia using group testing in real-world settings}\label{sec:application}

\textit{Chlamydia trachomatis} is the most commonly reported bacterial sexually transmitted infection (STI) in the United States \citep{mohseni:2023:chlamydia} and many other countries \citep{aseffa:1998:chlamydia}. Because infection is often asymptomatic, population-level screening is essential for early detection and treatment, thereby reducing the risk of serious complications such as pelvic inflammatory disease and infertility. Group testing is particularly useful in this setting because, when prevalence is low, pooled testing can substantially reduce the cost of large-scale screening. 

In practice, however, group testing for Chlamydia must accommodate several real-world complications. Diagnostic assays are imperfect, so both false positive and false negative results may occur. In addition, screening programs operate under budgetary and logistical constraints, so the cost of testing cannot be ignored. These practical considerations should therefore be incorporated directly into the design stage rather than treated only at the analysis stage.

Our goal is to construct a group testing design that yields precise estimation of the key parameters governing the screening process. In particular, we seek efficient estimation of the prevalence, denoted by \(p_0\), together with the sensitivity and specificity of the assay when tests are imperfect. We represent a design by
\[
\xi =
\begin{pmatrix}
x_1 & \cdots & x_k\\
w_1 & \cdots & w_k
\end{pmatrix},
\]
where \(x_1,\dots,x_k\) are the selected group sizes satisfying
\[
1 \le x_1 < \cdots < x_k \le M < \infty,
\]
and \(w_1,\dots,w_k\) are the corresponding design weights with \(w_\ell \ge 0\) for \(\ell=1,\dots,k\) and \(\sum_{\ell=1}^k w_\ell = 1\). Here, \(M\) denotes the largest allowable group size, so the design space is \(S=\{1,2,\ldots,M\}\). Following \citet{huang:2020:cost-GT}, we consider \(M=150\), and also examine a smaller setting with \(M=61\).

The design problem is driven by three practical ingredients: the prevalence to be estimated, the imperfect nature of the assay, and the cost of conducting the study.

\textbf{Prevalence.}
A primary objective of many group testing studies is to estimate disease prevalence as accurately as possible. In the simplest setting, prevalence can be estimated by maximum likelihood under a binomial model with perfect tests. In Chlamydia screening, however, this idealization is often unrealistic because the assay may produce both false positives and false negatives. Consequently, the design problem must account not only for prevalence, but also for the operating characteristics of the test itself.

\textbf{Imperfect tests.}
Let \(\btheta = (p_0,p_1,p_2),\) where \(p_0\) is the prevalence, \(p_1\) is the sensitivity, and \(p_2\) is the specificity. Following \citet{huang:2020:cost-GT}, we assume that false negatives and false positives occur randomly with probabilities \(1-p_1\) and \(1-p_2\), respectively, where \(p_0 \in (0,1)\) and \(p_1,p_2 \in (0.5,1]\). Under this model, the probability that a pooled test of size \(x\) yields a positive result is
\begin{equation}\label{eq-pi}
\pi(x) := \pi(x \mid \btheta)
= p_1 - (p_1+p_2-1)(1-p_0)^x.
\end{equation}
This probability includes both true and false positives and therefore reflects the actual operating characteristics of the assay.

For the Chlamydia Infertility Prevention Project (IPP), funded by the Centers for Disease Control and Prevention and analyzed in \citet{mcmahan:2012:informative}, the parameter values are
\[
\btheta_0 = (0.07,0.93,0.96),
\]
where the estimated prevalence is \(0.07\), the sensitivity is \(0.93\), and the specificity is \(0.96\). We use these values as the nominal parameter values in the design calculations in Section~\ref{sec:results}. Following \citet{huang:2020:cost-GT} and \citet{huangejs} , we use maximum group sizes of \(150\) and \(61\), corresponding to the large and moderate upper bounds used in \citet{huang:2020:cost-GT} and \citet{huangejs}, respectively.

\textbf{Balancing enrolment and assay costs.}
In addition to imperfect testing, practical group testing studies must account for cost. Let \(C_0\) denote the total budget for the experiment. Following \citet{huang:2020:cost-GT}, the cost of conducting one trial at group size \(x\) is modelled as
\[
c_0(x)=q_0+q_1x=(q_0+q_1)(1-q+qx),
\]
where \(q_0\) is the cost of performing an assay, \(q_1\) is the cost of enrolling one subject, and
\[
q=\frac{q_1}{q_0+q_1}.
\]
It is convenient to work with the standardized cost function
\[
c(x)=1-q+qx=\frac{c_0(x)}{q_0+q_1},
\qquad
C=\frac{C_0}{q_0+q_1},
\]
where \(C\) is the standardized total budget. Here \(q_0,q_1 \ge 0\) and \(q_0+q_1>0\). When \(q=0\), the cost does not depend on the group size, so \(c(x)=1\) for all \(x\), reducing the problem to the setting studied in \citet{huang:2017:design-group}. This case is appropriate when the dominant expense is the assay itself and the additional effort required to form larger pools is negligible. By contrast, when \(q>0\), larger group sizes incur additional enrolment or pooling costs and these must be incorporated into the design.

Taken together, prevalence estimation, imperfect assay performance, and cost constraints define the practical design problem for Chlamydia screening. The objective is to identify group sizes and allocation weights that yield efficient estimation of \(\btheta\) while respecting the operational realities of large-scale screening programs.

This perspective also clarifies why different design criteria may be appropriate for different scientific goals. As noted by \citet{li:2017-correlated-GT}, ``\emph{if one is only interested in estimating
the prevalence of a disease, retesting on subjects in the positive groups is not
necessary, since the probability of a group being positive is a monotone function of the probability of an individual being positive}." Accordingly, \(D_s\)-optimal designs prioritize prevalence estimation, \(c\)-optimal designs can target specific combinations of parameters, and \(D\)- and \(A\)-optimal designs seek efficient estimation of all components of \(\btheta\). We formalize these optimal design problems in the next section.

\section{General methodology: optimal design strategies for group testing experiments}\label{sec:OD}

The practical considerations discussed in Section~\ref{sec:application} naturally lead to search for an optimal design. This section presents general methodology for constructing a group testing experiment using the Chlamydia screening setting as an example. The aim is to choose group sizes and allocate resources across them so that the key parameters in the testing process can be estimated as precisely as possible, while respecting the operational constraints of the study. Because assay performance and cost both depend on the design, the resulting problem must be formulated in a way that jointly accounts for prevalence, imperfect tests, and budget.

\citet{huang:2017:design-group} used optimal design theory to construct group testing designs for a fixed sample size when the costs of testing and subject enrolment were the same across all group sizes. \citet{huang:2020:cost-GT} extended this framework to the more realistic setting in which cost depends on group size. Since the latter contains the former as a special case, we adopt the model in \citet{huang:2020:cost-GT} and build on it to develop a broader class of group testing designs for the Chlamydia problem.

Suppose data are collected under a design \(\xi\), and let \(\hat{\btheta}\) denote the maximum likelihood estimator of \(\btheta\). The asymptotic covariance matrix of \(\hat{\btheta}\) is proportional to the inverse of the Fisher information matrix \(\bI(\xi,\btheta)\). Accordingly, optimal design construction reduces to choosing \(\xi\) so as to optimize a suitable scalar function of \(\bI(\xi,\btheta)\).

For a design \(\xi\) with \(n_i\) trials at group size \(x_i\), \(i=1,\ldots,k\), the information matrix is
\begin{equation}
\bI(\xi,\btheta)
=
\sum_{i=1}^k
w_i \lambda(x_i)\bff(x_i)\bff^\top(x_i),
\label{eq:InfoGroup}
\end{equation}
where
\[
w_i = \frac{n_i c(x_i)}{C},
\qquad
C=\sum_{j=1}^k n_j c(x_j),
\qquad
\lambda(x)=\left[c(x)\pi(x)\{1-\pi(x)\}\right]^{-1},
\]
and
\[
\bff(x)=
\left(
x(p_1+p_2-1)(1-p_0)^{x-1},\,
1-(1-p_0)^x,\,
-(1-p_0)^x
\right).
\]
Here, \(\bff(x)\) is the derivative of the pooled-test positivity probability \(\pi(x)\) with respect to \(\btheta\), and therefore describes how the response probability changes with the underlying parameters. The weights \(w_i\) represent the proportion of total resources allocated to group size \(x_i\), and satisfy \(\sum_{i=1}^k w_i=1\).

When \(q=0\), we have \(c(x_i)=1\) for all \(i\), so that \(C=\sum_{j=1}^k n_j=n\), the total sample size, and \(w_i\) reduces to the proportion of observations assigned to group size \(x_i\); see \citet{huang:2020:cost-GT} for details. Thus, \eqref{eq:InfoGroup} provides a unified formulation covering both the constant-cost and group-size-dependent cost settings.

Following standard optimal design theory, we assess the quality of a design through its information matrix and define design criteria as scalar functions of \(\bI(\xi,\btheta)\). Single objective criteria are appropriate when the study has one primary inferential goal, whereas multi-objective criteria are useful when several aspects of \textcolor{black}{of the estimation problem} must be balanced simultaneously. We consider both settings below.

\subsection{Single objective design problems}

Many familiar optimality criteria arise as special cases of the general problem
\begin{equation}\label{eq:ProbOne}
\min_{\xi \in \Xi_S} \phi\!\left\{ \bI^{-1}(\xi,\btheta_0) \right\},
\end{equation}
where \(\bI(\xi,\btheta_0)\) is defined in \eqref{eq:InfoGroup}, \(\btheta_0\) is a nominal value of \(\btheta\), \(\Xi_S\) is the set of all designs on the design space \(S\), and \(\phi\) is a scalar criterion function. For example, \(\phi(\cdot)=\det(\cdot)\) corresponds to \(D\)-optimality, while \(\phi(\cdot)=\tr(\cdot)\) corresponds to \(A\)-optimality.

Because the information matrix depends on \(\btheta\), the solution \(\xi^*\) to \eqref{eq:ProbOne} depends on \(\btheta_0\) and is therefore a \emph{locally optimal design}. In this paper, we focus on locally optimal designs, which are widely used in practice and often serve as building blocks for more elaborate robust or sequential procedures. In applications, the nominal value \(\btheta_0\) is typically obtained from previous studies or reliable pilot information.

Solving \eqref{eq:ProbOne} requires determining both the support points \(x_i\) and their corresponding weights \(w_i\). In general, closed-form solutions are difficult to obtain, especially when the design depends on the parameters \(p_0,p_1,p_2\), the cost parameter \(q\), and the maximum group size \(M\). For the Chlamydia problem, \citet{huang:2017:design-group} and \citet{huang:2020:cost-GT} derived \(D\)- and \(D_s\)-optimal designs. Here we extend this line of work by also constructing \(A\)-, \(c\)- and \(E\)-optimal designs.

\textbf{Equivalent formulation on a discrete design space.} Because the design space \(S=\{1,\ldots,M\}\) is discrete, the design problem can be reformulated directly in terms of the weights assigned to the \(M\) possible group sizes. Let \(\xi_S\) denote a design on \(S\) with weights \(w_1,\ldots,w_M\), where \(w_j\ge 0\) for \(j=1,\ldots,M\) and \(\sum_{j=1}^M w_j=1\). Write \(\bw=(w_1,\ldots,w_M)\), and note that \(w_j=0\) simply means that group size \(j\) is not a support point of the design.

In this representation, the information matrix becomes
\[
\bI(\bw,\btheta)
:=\bI(\xi_S,\btheta)
=
\sum_{i=1}^M
w_i \lambda(x_i)\bff(x_i)\bff^\top(x_i),
\qquad x_i=i,\quad i=1,\ldots,M.
\]
Define
\[
\Delta_M
=
\left\{
\bw\in\mathbb{R}^M:
w_j\ge 0,\ j=1,\ldots,M,\ \sum_{j=1}^M w_j=1
\right\}.
\]
Then \eqref{eq:ProbOne} is equivalent to
\begin{equation}\label{eq:ProbTwo}
\min_{\bw\in\Delta_M}
\phi\!\left\{ \bI^{-1}(\bw,\btheta_0) \right\}.
\end{equation}

This reformulation is useful because \(\bI(\bw,\btheta)\) is linear in \(\bw\), and for many commonly used optimality criteria, the objective function \(\phi\{ \bI^{-1}(\bw,\btheta_0) \}\) is convex in \(\bw\). Since \(\Delta_M\) is also a convex set, \eqref{eq:ProbTwo} is a convex optimization problem for a broad class of criteria. Moreover, because \(M\) is moderate in practice, numerical optimization on the discrete design space is computationally feasible and yields optimal designs efficiently. We solve \eqref{eq:ProbTwo} using CVX \citep{grant:2020:cvx-guide}. Let \(\bw^*_\phi\) denote the solution to \eqref{eq:ProbTwo}   under the criterion $\phi$ on $S$.

For the Chlamydia application, we  {construct a variety of optimal designs  beyond those typically used in group-testing screening experiments, which are based on $D$-optimality. Specifically, we consider the following optimality criteria formulated in terms of a convex functional of the information matrix:
\begin{equation}\label{eq:loss}
\begin{alignedat}{2}
    &\phi_D(\bw)     = \det\!\left\{ \bI^{-1}(\bw,\btheta_0) \right\}, \qquad
    &\phi_{D_s}(\bw) = \bc_1^\top \bI^{-1}(\bw,\btheta_0) \bc_1, \\
    &\phi_A(\bw)     = \tr\!\left\{ \bI^{-1}(\bw,\btheta_0) \right\}, \qquad
    &\phi_c(\bw)     = \bc^\top \bI^{-1}(\bw,\btheta_0) \bc,\\
    &\phi_E(\bw)     = -\lambda_{\min}\left\{ \bI(\bw, \btheta_0) \right\},
\end{alignedat}
\end{equation}
where \(\bc_1=(1,0,0)\) and \(\bc\in\mathbb{R}^3\). In this setting, the \(D_s\)-criterion focuses solely on prevalence estimation, whereas the \(c\)-criterion targets estimation of the linear combination \(\bc^\top \btheta\), whose variance is \(\bc^\top \bI^{-1}(\bw,\btheta_0)\bc\). The \(A\)- and \(D\)-criteria, by contrast, assess overall precision for all components of \(\btheta\). All criteria in \eqref{eq:loss} are convex in \(\bw\) except \(\phi_D\); see \citet{wong:2019:cvx}. For computation under \(D\)-optimality, we therefore minimize \(\log\{\phi_D(\bw)\}\), which is also convex in \(\bw\).

To compare designs across criteria, we use the corresponding efficiency measures. For a design with a weight vector $\bf{w}$ on $S$, the \(D\)-, \(A\)-, \(c\) and \(E\)-efficiencies are
\begin{equation}\label{eq:eff}
\begin{aligned}
&\eff_{D}(\bw) =  \left\{ \frac{\phi_D(\bw_D^*)}
{\phi_D(\bw)} \right\}^{1/3},\qquad 
&\eff_{A}(\bw) =    \frac{\phi_A(\bw_A^*)}
{\phi_A(\bw)},\\
&\eff_{c}(\bw) =     \frac{\phi_c(\bw_c^*)}
{\phi_c(\bw)},
&\eff_{E}(\bw) = \frac{\phi_E(\bw)}
{\phi_E(\bw_E^*)},
\end{aligned}
\end{equation}
% \begin{equation}\label{eq:eff}
% \eff_D(\bw)
% =
% \left\{
% \frac{\phi_D(\bw_D^*)}{\phi_D(\bw)}
% \right\}^{1/3},
% \qquad
% \eff_A(\bw)
% =
% \frac{\phi_A(\bw_A^*)}{\phi_A(\bw)},
% \qquad
% \eff_c(\bw)
% =
% \frac{\phi_c(\bw_c^*)}{\phi_c(\bw)}, \eff_{E}(\bw) = \frac{\phi_E(\bw)}
% {\phi_E(\bw_E^*)},
% \end{equation}
where \(\bw_D^*, \bw_A^*\), \(\bw_c^*\) and \(\bw_E^*\) are the solutions to \eqref{eq:ProbTwo} under the corresponding single-objective criteria. The \(D_s\)-efficiency is a special case of \(\eff_c(\bw)\) obtained by taking \(\bc=\bc_1\).

\subsection{Multiple-objective optimal designs}\label{sec:moj}

In many practical studies, a single inferential target is insufficient. Although prevalence estimation is often central in group testing, investigators may also care about sensitivity, specificity, or particular linear combinations of these parameters. Moreover, these quantities may matter differently to different stakeholders, such as laboratory managers, public-health decision makers, and study investigators.  \textcolor{black}{In situations where group testing has heterogeneous risks, as described in \cite{econ}, one could also attach different importance  in the various risk group and willing to have different inferential accuracy levels.}  A design chosen to optimize only one target may therefore perform poorly on others and may be unacceptable in practice. These considerations motivate   multiple objective optimal designs.

Multiple-objective optimal design is well established in the broader design literature; see, for example, \citet{wong:2023:CVX} and \citet{gao:2025:multiple-objective}, who used maximin criteria to construct designs with good performance across several objectives. See also \citet{adaptive}, who employed multiple-objective designs to balance efficiency and ethical considerations in response-adaptive survival trials. We adopt this perspective here and construct maximin designs for the Chlamydia group testing problem.

\noindent\textbf{Maximin design criterion.} Let \(\Phi_1(\bw),\ldots,\Phi_K(\bw)\) denote the \(K\) criterion functions corresponding to the chosen objectives. These may be any combination of \(D\)-, \(A\)-, \(c\) and \(E\)-optimality criteria. Since \(D_s\)-optimality is a special case of \(c\)-optimality, it is included by choosing \(\bc=\bc_1\), and more than one \(c\)-criterion may be considered simultaneously if different linear combinations of \(\btheta\) are of interest.

Specifically, define
\[
\Phi_j(\bw)=
\begin{cases}
\log\{\phi_D(\bw)\}, & \text{if the \(j\)th criterion is \(D\)-optimality},\\[1mm]
\phi_A(\bw), & \text{if the \(j\)th criterion is \(A\)-optimality},\\[1mm]
\phi_c(\bw), & \text{if the \(j\)th criterion is \(c\)-optimality},\\[1mm]
\phi_E(\bw), & \text{if the \(j\)th criterion is \(E\)-optimality},
\end{cases}
\qquad j=1,\ldots,K,
\]
where \(\phi_D(\bw)\), \(\phi_A(\bw)\),  \(\phi_c(\bw)\) and \(\phi_E(\bw)\) are given in \eqref{eq:loss}. Let \(\eff_1(\bw),\ldots,\eff_K(\bw)\) denote the corresponding efficiency measures. The formulas in \eqref{eq:eff} provide these efficiencies for the criteria considered here.

A natural way to obtain protection across several objectives is to maximize the minimum efficiency. This leads to the maximin design problem
\begin{equation}\label{eq:ProblemThree}
\max_{\bw\in\Delta_M}
\min\left\{
\eff_1(\bw),\ldots,\eff_K(\bw)
\right\}.
\end{equation}
Directly solving \eqref{eq:ProblemThree} is difficult because of the nested maximization and minimization. However, \citet{gao:2025:multiple-objective} showed that this problem can be reformulated as a convex optimization problem. In particular, \eqref{eq:ProblemThree} is equivalent to
\begin{equation}\label{eq:ProblemFour}
\min_{\bw\in\Delta_M,\ t\ge 0} t
\quad
\text{subject to}
\quad
\Phi_j(\bw)\le h_j(t),\qquad j=1,\ldots,K,
\end{equation}
where
\begin{equation}\label{eq:Functionh}
 h_j(t)=
 \begin{cases}
 \log\{\phi_D(\bw_D^*)\}+3\log(t), & \text{if the \(j\)th criterion is \(D\)-optimality},\\[1mm]
 t\,\phi_A(\bw_A^*), & \text{if the \(j\)th criterion is \(A\)-optimality},\\[1mm]
 t\,\phi_c(\bw_c^*), & \text{if the \(j\)th criterion is \(c\)-optimality},\\
 \phi_E(\bw_E^*)/t, &\text{if the \(j\)th criterion is \(E\)-optimality},
 \end{cases}
\end{equation}
and \(\bw_D^*\), \(\bw_A^*\), \(\bw_c^*\)  and \(\bw_E^*\) are the corresponding single objective optimal designs.

The introduction of the auxiliary variable \(t\) is the key step in the reformulation. The constraints in \eqref{eq:ProblemFour} imply that
\[
\eff_j(\bw)\ge 1/t,\qquad j=1,\ldots,K,
\]
so minimizing \(t\) is equivalent to maximizing the minimum efficiency across all objectives. Let \(\bw^{**}\) and \(t^*\) be the solutions to \eqref{eq:ProblemFour}. Then \(\bw^{**}\) solves \eqref{eq:ProblemThree}, \(\xi_S(\bw^{**})\) is the maximin design, and
\[
\min \left\{
\eff_1(\bw^{**}),\ldots,\eff_K(\bw^{**})
\right\}
=
1/t^*.
\]
Although we focus on combinations of \(D\)-, \(A\)-, \(c\)- and \(E\)-criteria, the same strategy can be applied to other optimality criteria, such as \(A_s\)-, \(I\)-, and \(L\)-optimality  discussed in \cite{berger-wong:2009} after the function  \(h_j(t)\) is suitably modified. The corresponding optimality conditions are provided in the Supplementary.

\subsection{Optimal exact designs}\label{sec:exact}

The designs obtained above are optimal approximate designs. In practice, however, implementation requires exact numbers of trials at each group size. This gives rise to the problem of constructing optimal exact designs (OEDs) from the corresponding optimal approximate designs (OADs). Unlike OADs, exact designs are not naturally expressed as convex optimization problems, and equivalence theorems are generally unavailable to verify their optimality directly. A common strategy is therefore to convert an OAD into an OED by means of a rounding procedure; see, for example, \citet{pukelsheim:92:rounding}.

For the Chlamydia group testing problem, we consider two practically relevant settings: (\textbf{I}) \(q=0\), where the total sample size \(n\) is fixed and not large, and (\textbf{II}) \(q>0\), where the available budget \(C\) is fixed. We develop two rounding algorithms for these two settings and show that they perform well relative to existing approaches, including the method in \citet{huang:2020:cost-GT}.

  For a given OAD \(\xi_S(\bw^*)\) defined on $S$ with the optimal weight vector $\bw^*$, the basic idea is to round each of the quantities \(n w_i^*\)s to an integer \(n_i\) subject to \(\sum_i n_i=n\). In practice, however, this step is not always straightforward, especially when costs depend on the group size or the sample size is small. We therefore provide explicit rounding procedures. 
We write \(\xi_{RA}(\cdot)\) for an exact design obtained by a rounding algorithm in the fixed-sample-size case when \(q=0\), and write \(\xi_{RAC}(\cdot)\) for an exact design obtained by a rounding algorithm with cost constraint in the case when \(q>0\). Here, \(RA\) stands for ``rounding algorithm'' and \(RAC\) for ``rounding algorithm with cost constraint''.
 Rounding Algorithm I applies to the general case \(q>0\), where the exact design is constrained by the total budget \(C\), whereas Rounding Algorithm II applies to the case \(q=0\), where the total sample size \(n\) is fixed. 
%We first present the rounding algorithm for the budget-constrained case \(q>0\). }

\noindent \underline{\textbf{Rounding Algorithm I}}
\begin{description}[leftmargin=0.1em, labelsep=0.1em, itemsep=0em, parsep=0em, topsep=0em]
\item[\textbf{Step I.}]
 ~Suppose there are $k$ support points in an approximate design \(\xi_S(\bw^*)\),
and denote them as
$x_1, \ldots, x_k$.
For \(i=1,\ldots,k\), let \(n_i'\) be the largest integer less than or equal to \(Cw_i^*/c(x_i)\), that is, the floor of \(Cw_i^*/c(x_i)\). If \(\sum_{i=1}^k n_i' c(x_i)=C\), go to Step III; otherwise proceed to Step II.

\item[\textbf{Step II.}]
~Let
\[
C_r = C-\sum_{i=1}^k n_i' c(x_i),
\]
and let \(m_i\) be the floor of \(C_r/c(x_i)\) for \(i=1,\ldots,k\). Here \(C_r\) is the remaining budget after the initial allocation. If \(m_i=0\) for all \(i=1,\ldots,k\), go to Step III. Otherwise, search for adjustments (integers) \(\Delta_i\) satisfying \(\Delta_i\le m_i\) and
\[
\sum_{i=1}^k \Delta_i c(x_i)\le C_r,
\]
and choose integers \(\Delta_1,\ldots,\Delta_k\) so that the objective function is minimized. The objective function is the one used to obtain \(\xi_S(\bw^*)\).
Replace \(n_i'\) by \(n_i'+\Delta_i\), \(i=1,\ldots,k\).

\item[\textbf{Step III.}]
~Set \(n_i=n_i'\), \(i=1,\ldots,k\). The resulting exact design is denoted by
\[
\xi_{RAC}(C; n_1,\ldots,n_k;\xi_S(\bw^*)),
\]
with \(n_1,\ldots,n_k\) replicates at group sizes \(x_1,\ldots,x_k\), respectively.
\end{description}

%\violet{When \(C\) is large, Rounding Algorithm I typically produces highly efficient exact designs, but its performance may deteriorate when \(C\) is small. To improve performance in such cases, we augment Step II with a local search over nearby candidate support points.}

%\violet{We next consider the case \(q=0\), where the total sample size \(n\) is fixed. Suppose that there are \(k\) support points, \(x_1, \ldots, x_k\), in \(\xi_S(\bw^*)\), with corresponding weights \(w_1^*, \ldots, w_k^*\).}

\noindent \underline{\textbf{Rounding Algorithm II}}
\begin{description}[leftmargin=0.5cm, labelsep=0.1em, itemsep=0em, 
  parsep=1em,
  topsep=0.5em]
\item  \textbf{Step I}: For \(i=1, \ldots, k\), let \(n_i^\prime\) be the largest integer that is less than or equal to
\(n \cdot w_i^*\), i.e., \(n_i^\prime\) is the floor of \(n \cdot w_i^*\). If \(\sum_{i=1}^k n_i^\prime =n\), go to  Step III; otherwise
go to  Step II.
\item  \textbf{Step II}: For \(i=1, \ldots, k\), let \(r_i=n \cdot w_i^* -n_i^\prime\) and \(m=n-\sum_{i=1}^k n_i^\prime\).
Order  \(r_1, \ldots, r_k\) from the largest to the smallest.  Suppose \(r_{i_1}, \ldots, r_{i_{m}}\)
are the largest \(m\) values. Then we change \(n_{i_j}^\prime=n_{i_j}^\prime+1\) for \(j=1, \ldots, m\).
\item  \textbf{Step III}: For \(i=1, \ldots, k\), let \(n_i=n_i^\prime\). Then the exact design, denoted as \(\xi_{RA}(n_1, \ldots, n_k; \xi_S(\bw^*))\), has replicates
\(n_1, \ldots, n_k\) at the \(k\) support points of \(\xi_S(\bw^*)\), respectively.   
\end{description}

Our experience is that when \(C\) is large, Rounding Algorithm I typically produces highly efficient exact designs. Similarly Rounding Algorithm II usually performs well when \(n\) is moderate or large. However, when \(C\) is small or \(n\) is very small, the performances of the two  algorithms can be improved by  modifying  Step II of 
each algorithm slightly. Specifically, we slightly modify Step II of both rounding algorithms by enlarging the local search around each support point \(x_i\) to include nearby candidate points \(x_i \pm 1\) and \(x_i \pm 2\) provided they remain inside the design space \(S\). The remaining observations or budget are then allocated by searching over these local candidates so as to improve the objective value. This modification is especially helpful when the sample size \(n\) or budget \(C\) is small, where direct rounding alone may lead to less efficient or even singular exact designs. Although the additional local search increases computational cost, it remains far more practical than a complete greedy search, which becomes infeasible even for moderate values of \(M\) and \(n\). In our numerical studies, the proposed procedures typically require only 1--5 seconds to compute each exact design reported in Tables~\ref{tbl:MOD_rounding_budget-2},  \ref{tbl:MOD_rounding_combined}, and three Tables in the Supplementary material. Additional remarks on both algorithms are also given in the Supplementary material.

\section{Optimal designs for the Chlamydia group testing problem}\label{sec:results}

This section presents different types of optimal group testing designs developed in the previous sections. The OEDs   are obtained from Algorithms I and II. To facilitate implementation and reproducibility, we have developed an \textsf{R} package, \texttt{gtDesign} \citep{gtDesign}, for computing the group testing designs studied in this paper. The package is available on CRAN, and the corresponding GitHub repository provides additional documentation and reproducible code. All results reported below can be reproduced using the package.

\subsection{Single-objective optimal designs}

We begin with optimal approximate designs (OADs) under single-objective criteria. As discussed in Section~\ref{sec:exact}, these approximate designs do not depend on the total budget \(C\), whereas the corresponding exact designs do. We consider several values of the cost parameter \(q\) and the maximum group sizes \(61\) and \(150\) used in the reference papers as mentioned in Section~\ref{sec:application}, and compare the resulting \(D\)- and \(D_s\)-optimal designs with those reported in \citet{huang:2017:design-group,huang:2020:cost-GT}. The \(A\)-, \(c\)- and \(E\)-optimal designs are new.

Table~\ref{tbl:1-2} reports the OADs for \(M=61\) and \(M=150\) with \(q=0,~0.2,~0.8\), using the nominal parameter values \(\btheta_0=(0.07,0.93,0.96)\). For the \(c\)-optimal designs, we take \(\bc=(0,1,1)\). In addition to the support points and weights, the table reports the efficiency measures defined in \eqref{eq:eff}, thereby allowing direct comparison of each design under competing criteria.

Several features are apparent. First, most single-objective optimal designs have three support points, although a few cases with \(q=0.8\) require four. Second, equal weights arise only for the \(D\)-optimal designs, except when \(M=61\) and \(q=0.8\). Third, the \(D\)- and \(D_s\)-optimal designs agree with those reported in \citet{huang:2017:design-group,huang:2020:cost-GT}, providing a useful validation of our implementation. Finally, all designs include group size 1 as a support point. When \(q=0\), the upper boundary point \(M\) also appears frequently as a support point, whereas this does not occur once group-size-dependent costs are introduced.

A key message from Table~\ref{tbl:1-2} is that a design that is optimal for one criterion may perform poorly under another. For example, when \(q=0.2\) and \(M=61\), the \(D\)-optimal design has efficiencies \(\eff_A=0.744\), \(\eff_{D_s}=0.756\), \(\eff_E = 0.568\) and \(\eff_c=0.536\). Similarly, the \(D_s\)-optimal design has \(\eff_D=0.809\), \(\eff_A=0.624\), \(\eff_E=0.556\) and \(\eff_c=0.437\). The \(c\)-optimal designs are particularly striking: although optimal for their target criterion, they perform very poorly under the competing criteria. These results make clear that single-objective optimality alone may be insufficient in applications where several inferential goals must be considered simultaneously, which motivates the multi-objective designs studied next.

\begin{table}[!ht]
\renewcommand{\arraystretch}{0.6}
\centering
\caption{OADs under various criteria and cost values with \(\bc=(0,1,1)\) and \(\btheta_0 = (0.07, 0.93, 0.96)\). }
\resizebox{\textwidth}{!}{%
\begin{tabular}{cccrrrrrrrr} 
\toprule
& &          &  Support & & & & & \\
\(M\) & Case & Criteria & Points & Weights~~~~~~~~~~ & \( {\phi}(\bI^{-1})\)  & \(\eff_D\) & \(\eff_A\) & \(\eff_{D_s}\) & \(\eff_c\) & \(\eff_E\)  \\ 
\midrule
\multirow{18}{*}{61}  
& \multirow{4}{*}{\(q=0\)}  
 & \(D\) &1 17 61 & 0.333 0.333 0.333 & 0.003 & 1 & 0.936 & 0.705 & 0.692 & 0.946 \\
& & \(A\) &1 16 61 & 0.416 0.213 0.371 & 0.706 & 0.961 & 1 & 0.489 & 0.817 & 0.987  \\
& & \(D_s\) &1 16 61 & 0.131 0.628 0.241 & 0.035 & 0.811 & 0.509 & 1 & 0.337  &   0.438 \\
& & \(c\) &1 56 57 & 0.521 0.180 0.299 & 0.405 & 0.068 & 0.001 & 0 & 1  & 0.001 \\
& & \(E\) &1   16   61 & 0.415    0.250    0.335 & -2.36 &  0.979   &   0.991&      0.559  &     0.779    &    1   \\
\cmidrule{2-11}
& \multirow{4}{*}{\(q=0.2\)} 
 & \(D\) &1 10 61 & 0.333 0.333 0.333 & 0.135 & 1 & 0.744 & 0.756 & 0.536 &  0.568 \\
& & \(A\) &1 10 61 & 0.205 0.185 0.610 & 3.189 & 0.855 & 1 & 0.506 & 0.823 & 0.941  \\
& & \(D_s\) &1 10 61 & 0.106 0.569 0.325 & 0.147 & 0.809 & 0.624 & 1 & 0.437 & 0.556  \\
& & \(c\) &1 56 57 & 0.238 0.285 0.477 & 1.939 & 0.048 & 0.001 & 0 & 1 & 0 \\
& & \(E\) &1   10   61 & 0.126    0.188    0.686
 & -0.480 & 0.76 &      0.946&      0.509        & 0.773 &        1   \\
\cmidrule{2-11}
& \multirow{4}{*}{\(q=0.8\)} 
 & \(D\) &1 7 8 61 & 0.333 0.029 0.304 0.333 & 1.436 & 1 & 0.632 & 0.734 & 0.448 & 0.49   \\
& & \(A\) &1 8 61 & 0.125 0.183 0.692 & 9.191 & 0.753 & 1 & 0.496 & 0.823 & 0.941 \\
& & \(D_s\) &1 7 61 & 0.095 0.573 0.332 & 0.409 & 0.787 & 0.601 & 1 & 0.416  & 0.495\\
& & \(c\) &1 56 57 & 0.139 0.322 0.539 & 5.696 & 0.039 & 0.001 & 0 & 1 & 0 \\
& & \(E\) &1    7   61 & 
    0.063    0.165    0.772 & -0.151 & 0.602 &     0.928  &     0.44 &      0.787 &       1 \\
\midrule
\multirow{18}{*}{150}  
& \multirow{4}{*}{\(q=0\)}  
 & \(D\) & 1 19 150 & 0.333 0.333 0.333 & 0.021 & 1 & 0.903 & 0.694 & 0.738 & 0.779\\
& & \(A\) & 1 20 150 & 0.458 0.194 0.347 & 0.562 & 0.941 & 1 & 0.434 & 0.875 & 0.927 \\
& & \(D_s\) & 1 17 150 & 0.128 0.635 0.237 & 0.028 & 0.803 & 0.448 & 1 & 0.341 &  0.323 \\
& & \(c\) & 1 56 57 & 0.521 0.180 0.299 & 0.405 & 0.060 & 0.001 & 0 & 1 & 0 \\
& & \(E\) &1   17  150 & 
    0.532    0.208    0.260 &  -3.23& 0.917&      0.958      &0.458       &0.826 &        1   \\
\cmidrule{2-11}
& \multirow{4}{*}{\(q=0.2\)}  
 & \(D\) & 1 10 67 & 0.333 0.333 0.333 & 0.133 & 1 & 0.727 & 0.739 & 0.538 & 0.509  \\
& & \(A\) & 1 11 73 & 0.207 0.169 0.624 & 3.045 & 0.835 & 1 & 0.471 & 0.807 & 0.904  \\
& & \(D_s\) & 1 10 81 & 0.104 0.555 0.341 & 0.138 & 0.798 & 0.607 & 1 & 0.426 &  0.502 \\
& & \(c\) & 1 56 57 & 0.238 0.285 0.477 & 1.939 & 0.048 & 0.001 & 0 & 1&0 \\
& & \(E\) &1   10   81
    & 0.131    0.133    0.735 & -0.551 &0.692    &  0.925 &      0.393    &   0.729        &1    \\
\cmidrule{2-11}
& \multirow{4}{*}{\(q=0.8\)}  
 & \(D\) & 1 8 65 & 0.333 0.333 0.333 & 1.427 & 1 & 0.618 & 0.717 & 0.449 & 0.445  \\
& & \(A\) & 1 8 70 71 & 0.126 0.169 0.492 0.212 & 8.886 & 0.739 & 1 & 0.468 & 0.807 & 0.902 \\
& & \(D_s\) & 1 8 78 & 0.080 0.559 0.362 & 0.390 & 0.747 & 0.599 & 1 & 0.419 & 0.467 \\
& & \(c\) & 1 56 57 & 0.139 0.322 0.539 & 5.696 & 0.039 & 0.001 & 0 & 1 & 0\\
& & \(E\) & 1    8   78 & 
    0.057    0.112    0.830 & -0.169 & 0.517  &    0.881   &   0.328&       0.718 &        1   \\
\bottomrule
\end{tabular}
}
\label{tbl:1-2}
\end{table}

\subsection{Multi-objective optimal designs}

% \subsubsection{Maximin designs and equivalence verification}

Some group testing problems have multiple objectives and they have to be properly incorporated at the onset by choice of a multiple-objective design.  There are several ways of finding multiple-objective optimal designs and one approach is to search for a maximin design, which seeks to maintain good performance across multiple criteria rather than optimizing one criterion in isolation. Specifically, we compute maximin designs under dual- and triple-objective combinations using the formulation in Problem~\eqref{eq:ProblemFour}.

 Table~\ref{tbl:multiobj_maximin_time} reports selected multiple-objective optimal designs for \(M=61\) and \(M=150\) and there are additional results in the Supplementary materials. The results show a patterns as follows. First, most maximin designs again have three support points, although a few require four. Second, the achieved minimum efficiencies remain high across the scenarios considered: the dual-objective \(D\)-\(A\) designs all satisfy \(1/t^*>0.90\), and the triple-objective \(D\)-\(A\)-\(D_s\) designs all satisfy \(1/t^*>0.81\). Third, every maximin design retains group size 1 as a support point. For \(M=150\), the larger support points tend to decrease as \(q\) increases, reflecting the growing penalty associated with large pools; for \(M=61\), the upper boundary point remains a support point throughout. Finally, as expected, the achieved minimum efficiency declines as \(q\) increases.

These results show that multi-objective designs can retain high efficiency across competing criteria while remaining structurally simple. From a computational standpoint, the convex reformulation is also practical because all designs in Table~\ref{tbl:multiobj_maximin_time} were obtained in approximately 1--2 seconds.  An additional table provided in the Supplementary material has additional numerical results for the approximate maximin designs. 

\begin{table}[!ht]
\renewcommand{\arraystretch}{0.6}
\centering
\caption{Selected maximin designs for the dual-objective \(D\)-\(A\), triple-objective \(D\)-\(A\)-\(D_s\) and quadruple-objective \(D\)-\(A\)-\(c\)-\(E\)-criteria, with \(M=61,150\) and \(\btheta_0=(0.07,0.93,0.96)\). Here, \(1/t^*\) denotes the attained minimum efficiency, and Time is the computation time in seconds.}
\resizebox{\textwidth}{!}{%
\begin{tabular}{l|c|ccl|ccl} 
\toprule
\multicolumn{2}{c|}{Case} & \multicolumn{3}{c|}{\(M=150\)} & \multicolumn{3}{c}{\(M=61\)} \\ 
\cmidrule(lr){1-2} \cmidrule(lr){3-5} \cmidrule(lr){6-8}
 & \(q\) & Maximin designs \(\xi_S(\bw^{**})\) & \(1/t^*\) & Time & Maximin designs \(\xi_S(\bw^{**})\) & \(1/t^*\) & Time \\ 
\midrule
\multirow{4}{*}{\(D\)-\(A\)}
& 0    & \(\begin{pmatrix} 1 & 19 & 150 \\ 0.409 & 0.251 & 0.339 \end{pmatrix}\) & 0.981 & 1.534 & \(\begin{pmatrix} 1 & 16 & 17 & 61 \\ 0.382 & 0.114 & 0.148 & 0.356 \end{pmatrix}\) & 0.987 & 1.442 \\
\cmidrule(lr){2-8}
& 0.2  & \(\begin{pmatrix} 1 & 10 & 69 & 70 \\ 0.259 & 0.240 & 0.176 & 0.325 \end{pmatrix}\) & 0.943 & 1.359 & \(\begin{pmatrix} 1 & 10 & 61 \\ 0.258 & 0.248 & 0.494 \end{pmatrix}\) & 0.949 & 1.427 \\
\cmidrule(lr){2-8}
& 0.8  & \(\begin{pmatrix} 1 & 8 & 67 & 68 \\ 0.216 & 0.236 & 0.108 & 0.440 \end{pmatrix}\) & 0.909 & 1.453 & \(\begin{pmatrix} 1 & 7 & 61 \\ 0.217 & 0.242 & 0.541 \end{pmatrix}\) & 0.915 & 1.316 \\
\midrule
\multirow{4}{*}{\(D\)-\(A\)-\(D_s\)}
& 0    & \(\begin{pmatrix} 1 & 17 & 150 \\ 0.311 & 0.420 & 0.269 \end{pmatrix}\) & 0.818 & 1.660 & \(\begin{pmatrix} 1 & 15 & 16 & 61 \\ 0.289 & 0.246 & 0.179 & 0.286 \end{pmatrix}\) & 0.842 & 1.626 \\
\cmidrule(lr){2-8}
& 0.2  & \(\begin{pmatrix} 1 & 10 & 75 \\ 0.158 & 0.364 & 0.478 \end{pmatrix}\) & 0.855 & 1.586 & \(\begin{pmatrix} 1 & 10 & 61 \\ 0.156 & 0.377 & 0.467 \end{pmatrix}\) & 0.864 & 1.504 \\
\cmidrule(lr){2-8}
& 0.8  & \(\begin{pmatrix} 1 & 8 & 71 \\ 0.122 & 0.366 & 0.512 \end{pmatrix}\) & 0.848 & 1.594 & \(\begin{pmatrix} 1 & 7 & 8 & 61 \\ 0.123 & 0.158 & 0.217 & 0.502 \end{pmatrix}\) & 0.856 & 1.547 \\
\midrule
\multirow{4}{*}{\(D\)-\(A\)-\(c\)-\(E\)}
& 0    & \(\begin{pmatrix}
  1 & 26 & 150 \\ 
  0.469  &  0.203  &  0.328
\end{pmatrix}\) & 0.909 & 2.236&\(\begin{pmatrix}
  1 & 18 & 19 & 61 \\
  0.455 & 0.130 & 0.015 & 0.401
\end{pmatrix}\) &  0.891 & 2.001\\
\cmidrule(lr){2-8}

& 0.2  &\(\begin{pmatrix}
  1  & 11&   65   & 66\\
  0.237&    0.156 &   0.260 & 0.348
\end{pmatrix}\) &  0.844 &  2.190& \(\begin{pmatrix}
   1  &11 &    61 \\
  0.256 &    0.150 &    0.595
\end{pmatrix}\) & 0.848 &2.082 \\ 
\cmidrule(lr){2-8}
& 0.8  & \(\begin{pmatrix}
   1&     8 &   63   &64 \\
  0.189  &  0.162 &   0.250    &0.400
\end{pmatrix}\)& 0.812& 1.970& \(\begin{pmatrix}
    1 &    8&   61\\
  0.198 &    0.156 &    0.646
\end{pmatrix}\) & 0.814 & 2.018\\
\bottomrule
\end{tabular}
}
\label{tbl:multiobj_maximin_time}
\end{table}

How do we know the OADs in Tables 1 and 2 are optimal? When the optimality criterion is a convex functional of the information matrices and we have a large sample, there is theory exemplified in Theorem~1 in Supplementary material to confirm the optimality of an approximate design.  The theory relies on convex analysis results and produce equivalence theorems which are widely mentioned in the design monographs mentioned above.  These theorems can be verified graphically by plotting a certain function across $S$, and observing whether the plot exhibits certain properties.  Technical details are discussed in the Supplementary material with illustrative plots for a dual-objective optimal design problem. Figure~\ref{fig:equivalence3} below is another illustration using the same ideas to confirm optimality of  a design under the $D$-$A$-$D_s$-optimality criteria.  The first three subplots (a)-(c) violations the if and only if conditions in the equivalence theorem for checking whether the three-objective design is optimal under the $D$-, $A$- and $D_s$-optimality criteria.  The subplot (d) shows the design meets the conditions of the equivalence theorem and thus confirms optimality of the three-objective design.

\begin{figure}[ht]
  \centering
  \subfloat[]{\includegraphics[width=.23\textwidth]{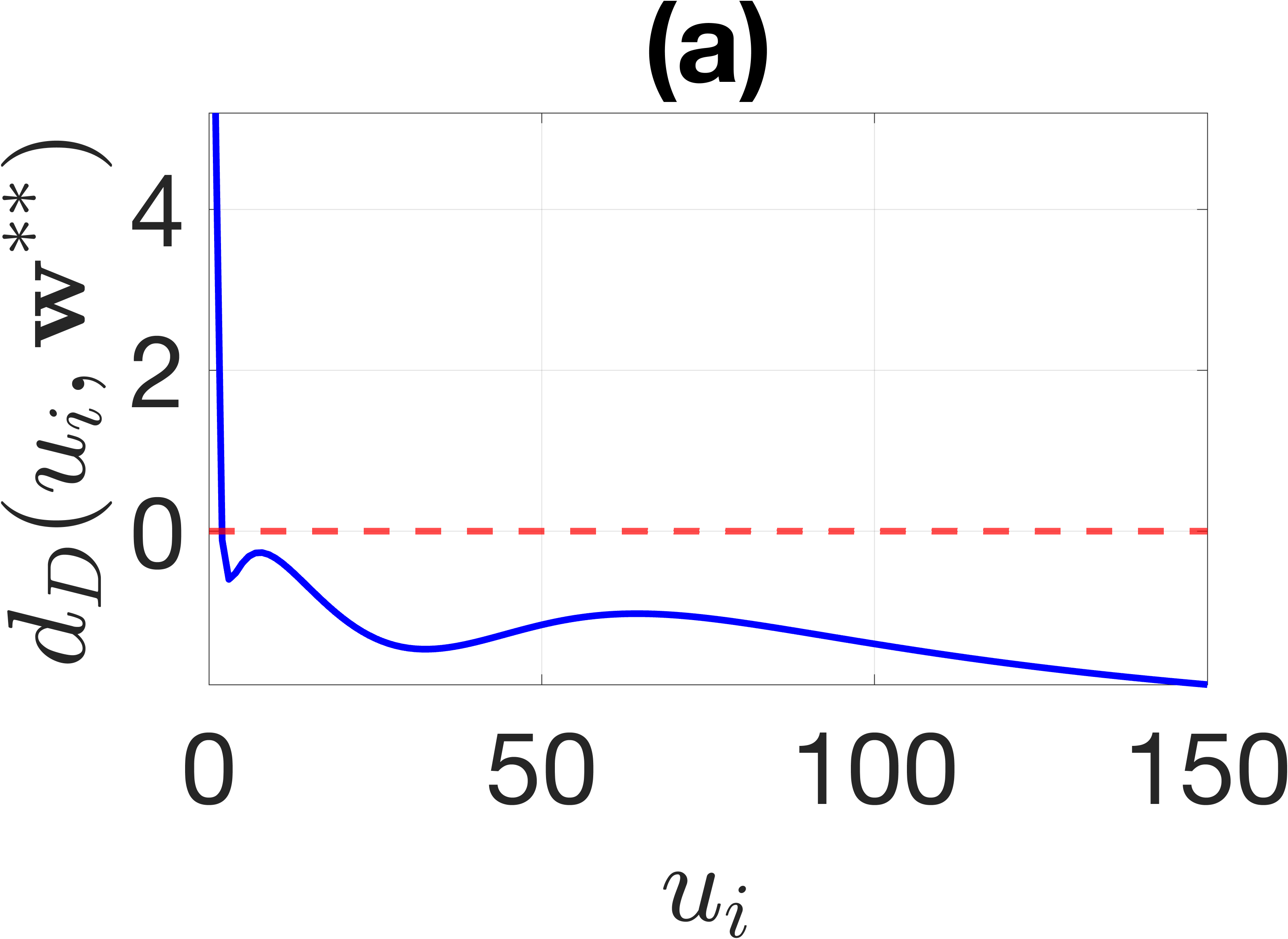}\label{fig:dD}}\hfill
  \subfloat[]{\includegraphics[width=.23\textwidth]{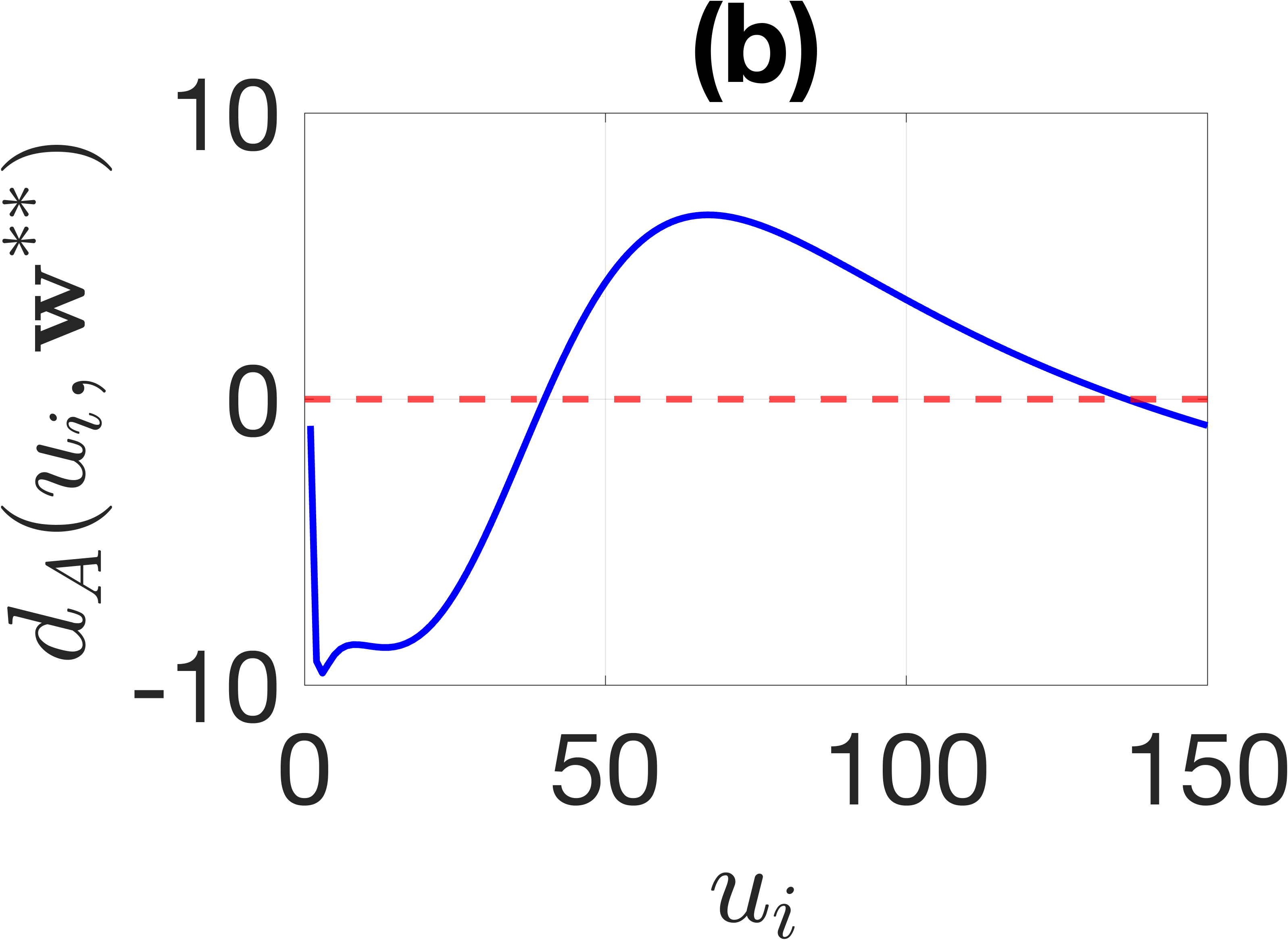}\label{fig:dA}}\hfill
  \subfloat[]{\includegraphics[width=.23\textwidth]{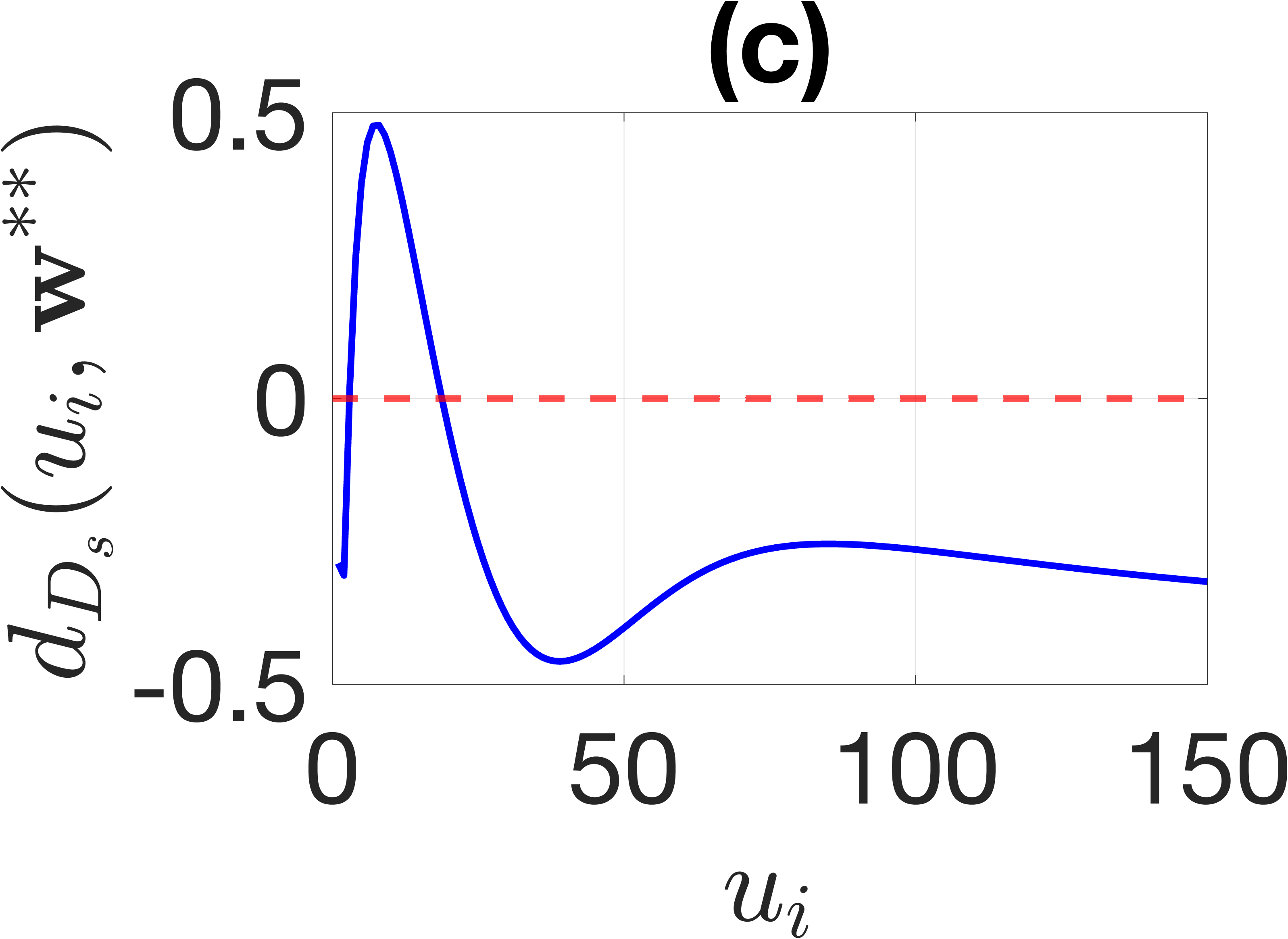}\label{fig:dDs}}\hfill
  \subfloat[]{\includegraphics[width=.23\textwidth]{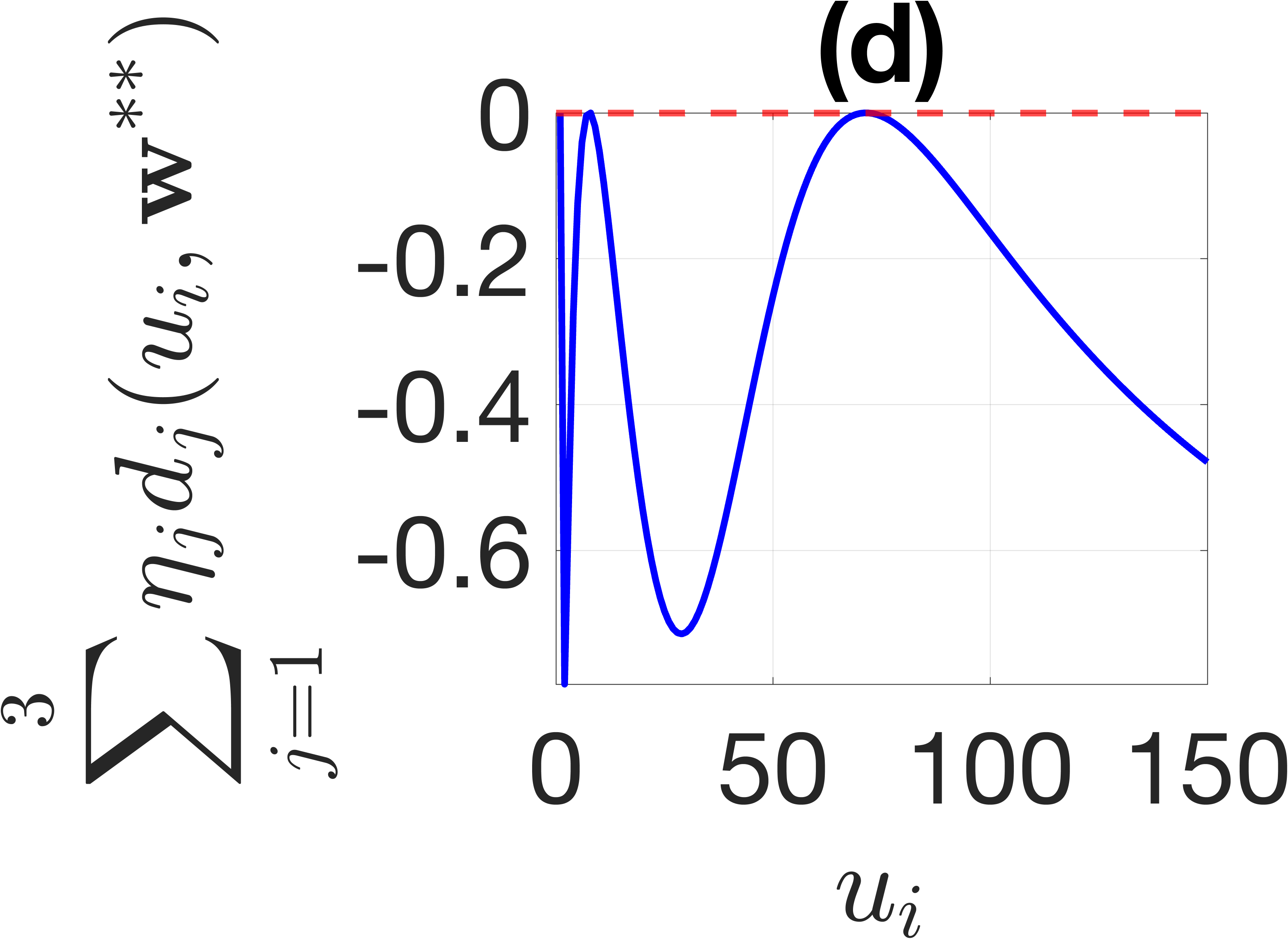}\label{fig:multi}}
  \caption{Verification of maximin design optimality for the \(D\)-\(A\)-\(D_s\) criterion with \(M=150\) and \(q=0.2\): (a) \(d_D(u_i,\bw^{**})\), (b) \(d_A(u_i,\bw^{**})\), (c) \(d_{D_s}(u_i,\bw^{**})\), and (d) the weighted sum \(\sum_{j=1}^3 \eta_j d_j(u_i,\bw^{**})\) with \((\eta_1,\eta_2,\eta_3)=(0,0.183,3.222)\) and \(t^*=1.170\).}
  \label{fig:equivalence3}
\end{figure}

\newpage
\subsection{Exact designs obtained by using Rounding Algorithms I and II} 
The approximate designs developed above provide the theoretical target, but implementation requires exact numbers of trials at each group size. We therefore apply Rounding Algorithms I and II to convert OADs into highly efficient exact designs under both fixed-sample-size (\(q=0\)) and fixed-budget (\(q>0\)) settings.

\medskip

\noindent\textbf{Single-objective exact designs.} Table~\ref{tbl:MOD_rounding_budget-2} displays exact designs for the single-objective criteria \(D\)-, \(A\)-, \(D_s\)-, \(c\)- and \(E\)-optimality with \(M=150\), \(q=0.2\), and several budget levels. All designs achieve efficiencies above \(0.96\), indicating that Rounding Algorithm I preserves the performance of the approximate designs extremely well. The table also reports the initial remaining budget \(C_r\), the adjustment vector \(\Delta\) from Step II of the algorithm, and the final remaining budget \(C_r^\prime\).

\begin{table}[!ht]
\renewcommand{\arraystretch}{0.4}
\centering
\caption{Single-objective exact designs obtained using Algorithm I with \(M=150\), \(\bc=(0,1,1)\), \(\btheta_0 = (0.07, 0.93, 0.96)\), and \(q=0.2\). \(\Delta\) indicates adjustments for remaining budget \(C_r\); \(C_r^\prime\) is final remaining budget.}
\resizebox{0.7\textwidth}{!}{%
\begin{tabular}{c|c c c c c c c} 
\toprule
Criterion & \(C\) & \(\xi_{RAC}\) & \(C_r\) & \(\Delta\) & \(C_r^\prime\) & \({\phi}(\bI^{-1})\) & \(\eff\) \\ 
\midrule
\multirow{3}{*}{\(D\)}    
 & 100  &\(\begin{pmatrix}
    1  &  10  &  11 &   67\\
  35  &  12 &    1  &   2
 \end{pmatrix}\) &7.8 & \(\begin{pmatrix}
   1 & 10 & 11 \\
   2 & 1 & 1 
 \end{pmatrix}\) & 0 &  0.135 & 0.994 \\
&  500 & \( \begin{pmatrix} 1 & 10 & 67 \\ 173 & 61 & 11 \end{pmatrix} \) & 12.6 & \( \begin{pmatrix} 1 & 10 \\ 7 & 2 \end{pmatrix} \) & 0 & 0.133 &  0.999 \\
& 10000 & \( \begin{pmatrix} 1 & 10 & 67 \\ 3334 & 1194 & 234 \end{pmatrix} \) & 12.2 & \( \begin{pmatrix} 1 & 10 \\ 1 & 4 \end{pmatrix} \) & 0 & 0.133 & 1 \\
\midrule 
\multirow{3}{*}{\(A\)}   
 & 100  &\(\begin{pmatrix}
    1 &   11 &   12 &   73\\
  20  &   5   &  1  &   4
 \end{pmatrix} \)&3.4  &\(\begin{pmatrix}
   12 \\
   1
 \end{pmatrix} \) & 0.2 &3.056  &0.997  \\
& 500 & \( \begin{pmatrix} 1 & 11 & 73 \\ 105 & 29 & 20 \end{pmatrix} \) & 5 & \( \begin{pmatrix} 1 & 11 \\ 2 & 1 \end{pmatrix} \) & 0 & 3.046 & 1 \\
&  10000 & \( \begin{pmatrix} 1 & 11 & 73\\ 2071 & 564 & 405 \end{pmatrix} \) & 7 & \( \begin{pmatrix} 1 & 11 \\ 4 & 1 \end{pmatrix} \) & 0 & 3.045& 1 \\
\midrule 
\multirow{3}{*}{\(D_s\)}    
 & 100  & \(\begin{pmatrix}
     1 &   10 &   81\\
  10   & 20   &  2
 \end{pmatrix}\)& 2.8 &  \(\begin{pmatrix}
   10 \\
   1
 \end{pmatrix}\)& 0 & 0.138 & 1 \\
&  500 & \( \begin{pmatrix} 1 & 10 & 81 \\ 52 & 99 & 10 \end{pmatrix} \) & 0.8 & None & 0.8 & 0.138 & 0.998 \\
&  10000 & \( \begin{pmatrix} 1 & 10 & 81 \\ 1042 & 1985 & 200 \end{pmatrix} \) & 11.2 & \( \begin{pmatrix} 10 \\ 4 \end{pmatrix} \) & 0 & 0.138 & 1 \\
\midrule 
\multirow{3}{*}{\(c\)}    
 & 100  &\( \begin{pmatrix}
   1 & 56 & 57 \\
   27 & 2 & 4 
 \end{pmatrix}\)& 16.4&  \(\begin{pmatrix}
   1 & 57 \\
   4 & 1 
 \end{pmatrix}\)& 0.2  & 1.967 &0.986  \\
& 500 & \( \begin{pmatrix} 1 & 56 & 57 \\ 124 & 12 & 19 \end{pmatrix} \) & 17.2 & \( \begin{pmatrix} 1 & 56 \\ 5 & 1 \end{pmatrix} \) & 0.2 & 1.941 & 0.999 \\
&  10000 & \( \begin{pmatrix} 1 & 56 & 57 \\ 2386 & 238 & 390 \end{pmatrix} \) & 16 & \( \begin{pmatrix} 1 & 56 \\ 4 & 1 \end{pmatrix} \) & 0 & 1.939 & 1 \\
\midrule
\multirow{6}{*}{\(E\)} 
 & 100  & \(\begin{pmatrix}
      1  &   9  &  10 &   81\\
    13  &   3   &  4 &    4
 \end{pmatrix} \)& 7.8 & \(\begin{pmatrix}
     9 \\
     3 
 \end{pmatrix} \)& 0 & -0.532 & 0.966\\
& 500 &\(\begin{pmatrix}
     1   &  9  &  10  &  81\\
    68  &   3 &   24  &  21
\end{pmatrix} \)& 13.6   & \(\begin{pmatrix}
    1 & 9 & 10 \\
    3 & 3 & 1
\end{pmatrix}\) & 0 & -0.547 &  0.994\\
&  10000 & \(\begin{pmatrix}
     1    &      10 &         81\\
        1312 &        480     &    432
\end{pmatrix}\)& 12.2 & \(\begin{pmatrix}
    1 & 10 \\
    1 & 4
\end{pmatrix}\)& 0 & -0.551 & 1\\
\bottomrule
\end{tabular}
 }
\label{tbl:MOD_rounding_budget-2}
\end{table}

\noindent\textbf{Multiple objective exact designs.}
Table~\ref{tbl:MOD_rounding_combined} reports representative multi-objective exact designs. For these designs, the minimum efficiency, denoted MinEff, is defined as \(\min\{\eff_1,\ldots,\eff_K\}\) across the objectives under consideration. The main finding is that the exact designs closely track their approximate counterparts: MinEff approaches \(1/t^*\), the efficiency of the corresponding OAD, as the sample size \(n\) or budget \(C\) increases. When \(n\) or \(C\) is small, the discrepancy is modest, typically around 0.018, and becomes negligible as more resources become available. This pattern is illustrated in Figure~\ref{fig:eff-MO-exact}.

\begin{table}[!ht]
\renewcommand{\arraystretch}{0.3}
\centering
\caption{Exact designs for \(M=61\) under multi-objective criteria with fixed sample sizes (\(q=0\)) obtained using Algorithm II or budgets (\(q=0.2\)) obtained using Algorithm I. MinEff represents \(\min\{\eff_1, \ldots, \eff_K\}\) at the exact design; \(1/t^*\) is from Table~\ref{tbl:multiobj_maximin_time}.}
\resizebox{0.8\textwidth}{!}{%
\begin{tabular}{cccccc|c} 
\toprule
Setting & Criterion & \(n\) or \(C\) & \(\xi_{RAC}\) & \(\Delta\) & MinEff & \(1/t^*\)\\ 
\midrule
\multirow{12}{*}{\(q=0\)} 
 & \multirow{3}{*}{\(D\)-\(A\)}  
 & \(n=10\)  & \(\begin{pmatrix} 1 & 15 & 16 & 17 & 61\\ 4 & 1& 1 & 1 & 3 \end{pmatrix}\) & \(\begin{pmatrix} 1 & 15 \\ 1 & 1 \end{pmatrix}\) & 0.959 & \multirow{3}{*}{0.987} \\
 && \(n=25\)  & \(\begin{pmatrix} 1 & 16 & 17 & 61\\ 10 & 2 & 4 & 9 \end{pmatrix}\) & \(\begin{pmatrix} 1 & 17 & 61\\ 1 & 1 &1  \end{pmatrix}\) & 0.977 & \\
 && \(n=50\)  & \(\begin{pmatrix} 1 & 16 & 17 & 61\\ 19  &   5 &     8 &   18 \end{pmatrix}\) & \(\begin{pmatrix} 17 & 61 \\ 1 & 1 \end{pmatrix}\) & 0.986 & \\
\cmidrule{2-7}
 & \multirow{3}{*}{\(D\)-\(A\)-\(D_s\)} 
 & \(n=10\)  & \(\begin{pmatrix} 1 & 15 & 16 & 61\\ 3 & 3 & 1 & 3 \end{pmatrix}\) & \(\begin{pmatrix} 1 & 15 & 61 \\ 1 & 1 & 1 \end{pmatrix}\) & 0.809 & \multirow{3}{*}{0.842} \\
 && \(n=25\)  & \(\begin{pmatrix} 1 & 15 & 16 & 17& 61\\ 7 &    6&     4 &    1  &    7 \end{pmatrix}\) & \(\begin{pmatrix} 17 \\ 1 \end{pmatrix}\) & 0.825 & \\
 && \(n=50\)  & \(\begin{pmatrix} 1 & 15 & 16 & 61\\ 14 & 13 & 8 & 15 \end{pmatrix}\) & \(\begin{pmatrix} 15 & 61 \\ 1 & 1 \end{pmatrix}\) & 0.838 & \\
\midrule
\multirow{8}{*}{\(q=0.2\)} 
 & \multirow{2}{*}{\(D\)-\(A\)}  
 & \(C=100\)  & \(\begin{pmatrix} 1 & 10 & 59 & 61\\ 26 & 8 &1 & 3 \end{pmatrix}\) & \(\begin{pmatrix} 1 & 59 \\ 1 & 1 \end{pmatrix}\) & 0.932 & \multirow{2}{*}{0.949} \\
 && \(C=500\)  & \(\begin{pmatrix} 1 & 10 & 60 & 61\\ 130 & 44 & 1 & 18 \end{pmatrix}\) & \(\begin{pmatrix} 1 & 60 \\ 1 & 1 \end{pmatrix}\) & 0.948 & \\
\cmidrule{2-7}
 & \multirow{2}{*}{\(D\)-\(A\)-\(D_s\)} 
 & \(C=100\)  & \(\begin{pmatrix} 1 & 10 & 61\\ 24 & 13 & 3 \end{pmatrix}\) & \(\begin{pmatrix} 1 \\ 9 \end{pmatrix}\) & 0.818 & \multirow{2}{*}{0.864} \\
 && \(C=500\)  & \(\begin{pmatrix} 1 & 10 & 61\\ 78 & 67 & 18 \end{pmatrix}\) & \(\begin{pmatrix} 61 \\ 1 \end{pmatrix}\) & 0.861 & \\
\bottomrule
\end{tabular}
}
\label{tbl:MOD_rounding_combined}
\end{table}

\begin{figure}[!ht]
  \centering
  \resizebox{\textwidth}{!}{%
  \subfloat[]{%
    \includegraphics[width=0.45\textwidth]{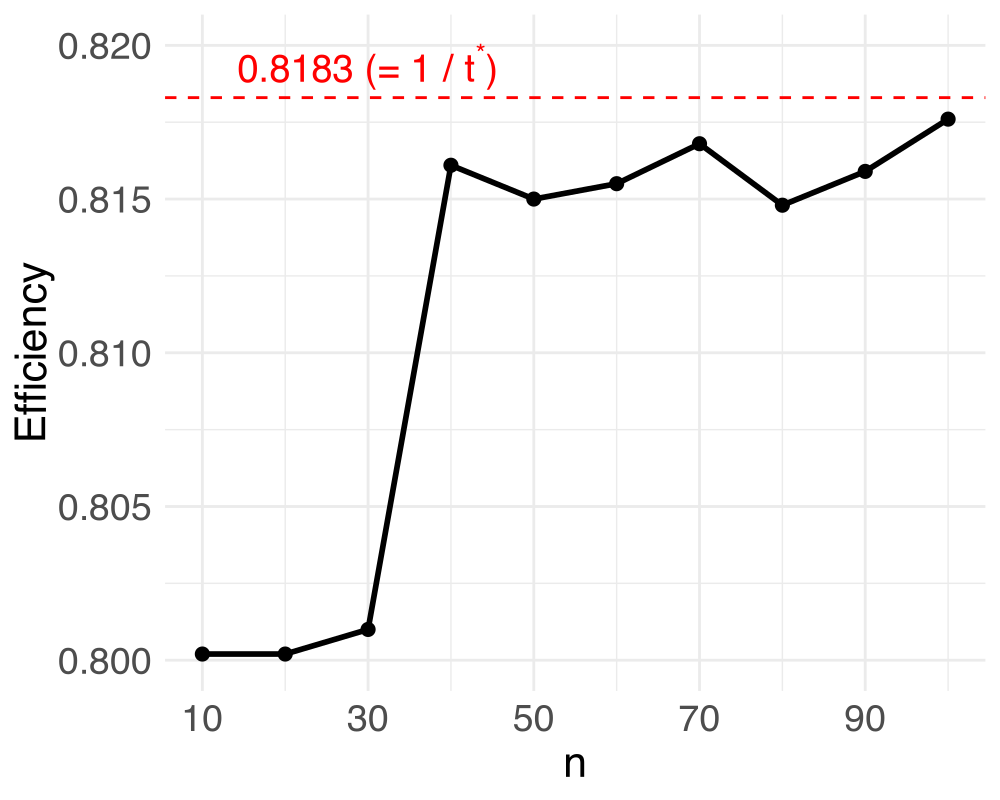}%
    \label{fig:mo_exact_eff1}}%
  \hfill
  \subfloat[]{%
    \includegraphics[width=0.45\textwidth]{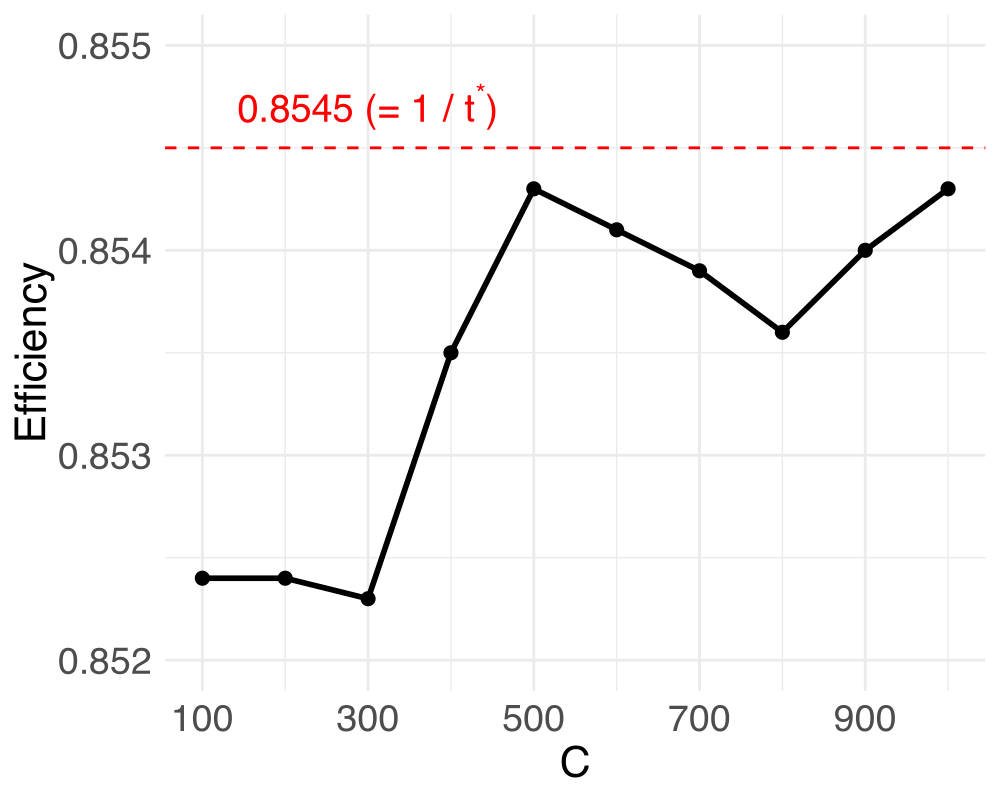}%
    \label{fig:mo_exact_eff2}}%
  }
  \caption{Efficiency of exact designs (MinEff) for multi-objective \(D\)-\(A\)-\(D_s\) designs with \(M=150\):  
  (a) MinEff versus sample size \(n\) for \(q=0\);  
  (b) MinEff versus budget \(C\) for \(q=0.2\). The dashed line shows \(1/t^*\) for the corresponding approximate design.}
  \label{fig:eff-MO-exact}
\end{figure}

Taken together, the exact-design results show that the proposed rounding algorithms are highly effective. Once the available budget or sample size is moderate, the resulting exact designs are uniformly efficient. In addition, exact designs often introduce one or more additional support points beyond those in the corresponding approximate design, especially when resources are limited. Finally, the remaining unspent budget after Step II is typically zero or negligible, and the computational burden remains modest, with run times of roughly 1--5 seconds per design. The preceding results show that the proposed designs improve the precision of parameter estimation under realistic cost and sample-size constraints. Since these estimated parameters are subsequently used as inputs in operational group-testing procedures, it is also of interest to examine how parameter misspecification may affect downstream testing efficiency.

\subsection{Downstream impact on Stage II assay burden}
\label{sec:stage2-downstream}

The preceding numerical results focus on Stage I, where the goal is to estimate
\(\btheta=(p_0,p_1,p_2)\) as precisely as possible under realistic cost and sample-size constraints. Since these estimates are subsequently used as inputs in operational group-testing procedures, it is also of interest to examine how parameter misspecification affects downstream testing efficiency.

To illustrate this point, we consider a Stage II screening programme with \(N=10{,}000\) subjects and candidate group sizes \(x\in\{1,2 \ldots,20\}\). Using the pooled-test positivity probability \(\pi(x)\) defined in \eqref{eq-pi}, we choose the Stage II group size by minimizing the total-testing criterion
\[
E_{\btheta}\{T(x)\}\approx \Big\lceil \frac{N}{x}\Big\rceil + N\,\pi(x).
\]
Thus, for each assumed parameter vector \(\tilde{\btheta}\), the Stage II choice is
\[
x^*(\tilde{\btheta})=\arg\min_{x\in\mathcal X}E_{\tilde{\btheta}}\{T(x)\}.
\]
We fix the true parameter vector at \(\btheta_0=(0.07,0.93,0.96)\), corresponding to the Chlamydia application in Section~\ref{sec:application}. Under the true parameter values, the oracle Stage II choice is \(x^*(\btheta_0)=5\), with an expected total of \(5108.374\) tests. We then evaluate a range of misspecified parameter scenarios and record the resulting Stage II choice \(x^*(\tilde{\btheta})\), together with the actual expected number of tests when that choice is evaluated under the true parameter vector \(\btheta_0\). An additional table in the supplementary material provides the misspecification details.

Table~\ref{tab:stage2_grouped} summarizes the main pattern. Some misspecified parameter values still lead to \(x=5\), in which case the regret is zero. Several mild misspecifications lead to \(x=4\), increasing the expected number of tests by about \(34\) tests (\(0.665\%\)). Stronger misspecifications lead to \(x=6\), increasing the expected number of tests by about \(100\) tests (\(1.966\%\)). More severe misspecifications lead to \(x=3\) or \(x=8\), increasing the expected number of tests by about \(367\) and \(461\), respectively, corresponding to \(7.182\%\) and \(9.031\%\) increases over the oracle choice. Figure~\ref{fig:stage2_regret} complements Table~\ref{tab:stage2_grouped} by visually displaying the magnitude of the extra expected tests (i.e., regret) across scenarios relative to the oracle benchmark. In particular, it shows that mild misspecification leads to only negligible regret, whereas strong overestimation and severe underestimation result in substantially larger losses.

These results show that the downstream consequences of parameter misspecification are not merely theoretical. Once the assumed parameter values are inaccurate enough to alter the Stage II choice of \(x\), the resulting decision produces a positive regret and unnecessary additional tests. This provides direct downstream support for the role of Stage I design: by improving the precision of parameter estimation, the proposed framework increases the chance that Stage II will recover the correct or nearly correct group size, thereby reducing unnecessary assay use. Additional details of this simulation are provided in Supplementary, including the full set of misspecified parameter scenarios and additional plots summarizing the resulting Stage II choices, regrets, and expected test burdens under the true parameter vector.

This point becomes even more relevant when the total screening budget is limited and must be allocated across both stages. In such settings, allocating more resources to Stage II necessarily leaves fewer resources for Stage I, which may in turn reduce the precision of the parameter estimates used downstream. Consequently, under a constrained total budget, Stage I design efficiency becomes more important rather than less important. When only a limited amount of first-stage information can be collected, an efficient design is essential for extracting as much information as possible from the available resources and for protecting the quality of the subsequent Stage II decision.

\begin{table}[ht]
\centering
\small
\caption{Downstream Stage II consequences of parameter misspecification under the true Chlamydia parameter vector \(\btheta_0=(0.07,0.93,0.96)\).}
\begin{tabularx}{\textwidth}{c c c c X}
\hline
\(x\) & Extra assays & Increase (\%) & Expected tests under \(\btheta_0\) & Misspecification pattern \\
\hline
5 & 0 & 0.000 & 5108.374 & \textbf{True parameter values} or near truth \\
4 & 33.963 & 0.665 & 5142.337 & Mild misspecification \\
6 & 100.413 & 1.966 & 5208.787 & Moderate underestimation \\
3 & 366.849 & 7.182 & 5475.223 & Strong overestimation \\
8 & 461.348 & 9.031 & 5569.722 & Severe underestimation \\
\hline
\end{tabularx}
\label{tab:stage2_grouped}
\end{table}

\begin{figure}[ht]
\centering
\includegraphics[width=0.75\textwidth]{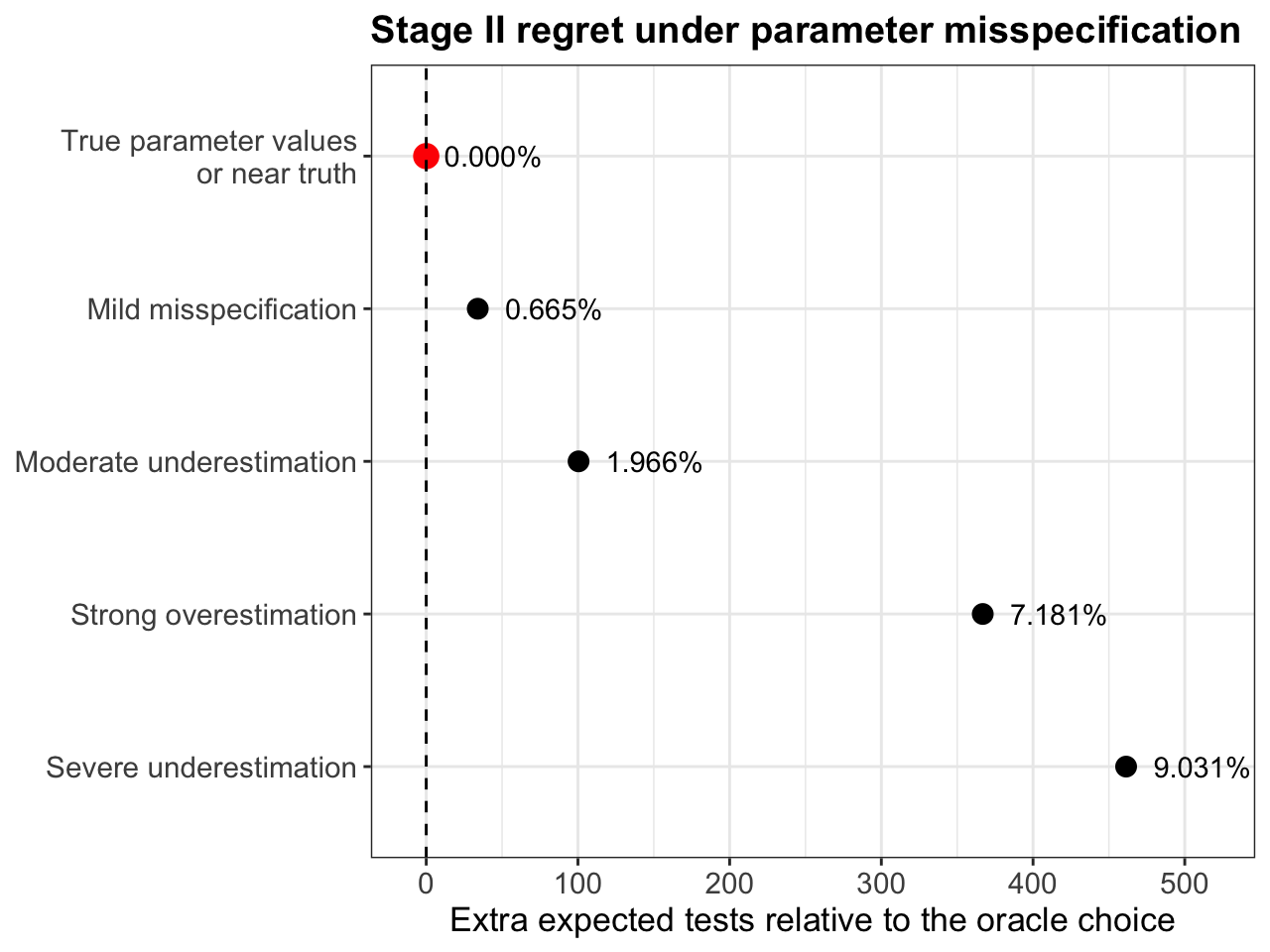}
\caption{Regret in Stage II under grouped misspecification scenarios, measured by the increase in expected tests relative to the oracle choice under the true parameter vector.}
\label{fig:stage2_regret}
\end{figure}

\section{Discussion and extensions}\label{sec:discussion}

The proposed framework is designed to address the types of design choices that arise in practical group testing studies, where investigators must allocate limited resources under cost constraints, balance competing inferential goals, and work within operational limitations. It covers both single- and multiple-objective design problems, which is especially important when no single inferential target is sufficient for decision-making. In such settings, compromise designs may be preferable to designs targeted to only one objective; see, for example, \citet{cookwong}, \citet{huangwong}, and \citet{adaptive}.

A further strength of the framework is that it allows maximin formulations to balance competing objectives in a principled and computationally tractable way. This is especially useful when a study requires protection against poor performance under different inferential priorities or model configurations. More broadly, the methodology is not tied to the Chlamydia application considered here, but is applicable whenever the information matrix can be expressed linearly in the design weights. The examples below illustrate how the proposed framework can be adapted to a broader class of practically relevant group testing problems.

\subsection{Exact designs under alternative operational constraints}

In practice, exact group testing designs are determined not only by statistical objectives, but also by operational constraints. Two common examples are limits on the total number of individuals that can be recruited and limits on the total number of assays or test kits available. Our framework can accommodate both, and the resulting designs may differ substantially even when the underlying disease and assay parameters remain unchanged.

To illustrate this point, we consider a second Chlamydia example based on the National Health and Nutrition Examination Survey (NHANES), as reported by \citet{datta:2007:chlamydia-data} and analyzed by \citet{li:2017-correlated-GT}. The corresponding parameter values are
\[
\btheta^{(C)}=(0.022,0.92,0.965),
\]
where \(p_0=0.022\) is the estimated prevalence, \(p_1=0.92\) the sensitivity, and \(p_2=0.965\) the specificity. Following \citet{li:2017-correlated-GT}, we set the maximum group size to \(M=15\) and consider the case \(q=0\), corresponding to a setting in which the assay cost dominates and the additional cost of forming larger pools is negligible.

This example is useful because it shows how the same inferential problem can lead to different exact designs depending on the type of resource constraint imposed. It also complements the analysis in \citet{li:2017-correlated-GT}, whose focus was on correlation modelling under a single \(D\)-optimality criterion. By contrast, our framework accommodates multiple criteria and exact constraints on either the number of individuals or the number of assays.

Table~\ref{tbl:app-approximate} reports the corresponding optimal approximate designs. All five criteria place support on the same three pool sizes, 1, 7, and 15, but the allocations differ noticeably. The \(D\)-optimal design assigns equal weight to all three support points, whereas the \(A\)-, \(E\)-, \(c\)-, and \(D_s\)-optimal designs place more emphasis on the intermediate pool size 7. Thus, even before discretization, the scientific objective materially affects the implied screening strategy. Notably, although the values of the loss function differ, the optimal designs under \(E\)-optimality and \(A\)-optimality are nearly identical in this setting.

\begin{table}[h]
\centering
\caption{Optimal approximate designs for Chlamydia with \(\btheta^{(C)}=(0.022,0.92,0.965)\), \(M=15\), and \(q=0\).}
\begin{tabular}{cccc}
\hline
Criterion & Pool size 1 & Pool size 7 & Pool size 15 \\
\hline
\(c\) & 0.155 & 0.519 & 0.326 \\
\(D\) & 0.333 & 0.333 & 0.333 \\
\(A,E\) & 0.159 & 0.517 & 0.324 \\
\(D_s\) & 0.173 & 0.526 & 0.301 \\
\hline
\end{tabular}
\label{tbl:app-approximate}
\end{table}

When the total number of individuals is fixed at \(m=6632\), matching the NHANES sample size, the corresponding exact designs are reported in Table~\ref{tbl:app-exact-individual}. The allocations differ appreciably across criteria. In particular, the discretized \(D\)-optimal design introduces an additional pool size 8 in order to satisfy the individual constraint exactly, whereas the other criteria retain three support points. The total number of assays ranges from 764 to 865, showing that the optimality criterion can have a substantial effect on resource use even when the total number of individuals is fixed.

\begin{table}[h]
\centering
\caption{Optimal exact designs for Chlamydia under the individual constraint \(m=6632\), with \(\btheta^{(C)}=(0.022,0.92,0.965)\) and \(q=0\). Numbers in parentheses indicate the number of tests at each pool size.}
\begin{tabular}{cccccccc}
\hline
Criterion & \multicolumn{5}{c}{Pool sizes and number of tests} & \# Tests & \# Individuals \\
\hline
\(D\) & 1 (288) & 7 (288) & 8 (1) & 15 (288) & & 865 & 6632 \\
\(A\), \(E\) & 1 (121) & 5 (1) & 7 (396) & 14 (1) & 15 (248)  & 767 & 6632 \\
\(c\) & 1 (118) & 7 (397) & 15 (249) & && 764 & 6632 \\
\(D_s\) & 1 (136) & 7 (418) & 15 (238) && & 792 & 6632 \\
\hline
\end{tabular}
\label{tbl:app-exact-individual}
\end{table}

By contrast, when the primary constraint is the number of available assays rather than the number of individuals, the exact designs look rather different. Table~\ref{tbl:app-num_tests} reports the designs under the assay-budget constraint \(C=500\). In this setting, all criteria retain the same support points as in the approximate designs, namely 1, 7, and 15, but the allocations again vary considerably. The \(D\)-optimal design distributes resources nearly equally, whereas the \(c\)-optimal design places the greatest emphasis on the intermediate pool size. The total number of individuals screened ranges from 3826 to 4336, illustrating that the nature of the operational constraint, individuals versus assays, can itself alter the practical design in a meaningful way.

\begin{table}[h]
\centering
\caption{Optimal exact designs for Chlamydia under the assay-budget constraint \(C=500\), with \(\btheta^{(C)}=(0.022,0.92,0.965)\) and \(q=0\). Numbers indicate the number of tests at each pool size.}
\begin{tabular}{ccccccc}
\hline
Criterion & Pool size 1 & Pool size 7 & Pool size 15 & \# Tests & \# Individuals \\
\hline
\(D\) & 167 & 167 & 166 & 500 & 3826 \\
\(A\), \(E\) & 80 & 258 & 162 & 500 & 4316 \\
\(c\) & 78 & 259 & 163 & 500 & 4336 \\
\(D_s\) & 86 & 263 & 151 & 500 & 4192 \\
\hline
\end{tabular}
\label{tbl:app-num_tests}
\end{table}

Taken together, these results show that exact design construction is sensitive not only to the choice of optimality criterion, but also to the type of resource constraint imposed. This reinforces an important practical point: in applied group testing, the design should be tailored jointly to the inferential objective and to the way resources are actually constrained.

\subsection{Group testing with two assays}

Another natural extension arises when more than one assay is available. This setting was studied by \citet{huangejs}, who considered a regular assay that is less expensive but subject to error, and a gold-standard assay that is perfectly accurate but more costly. In that framework, a group testing design may involve three procedures: testing a pooled sample using only the regular assay (Reg), only the gold-standard assay (Gold), or both assays (Both). Although the resulting information matrices differ from those used in the present paper, our framework extends readily after suitable modification of the information matrix and the associated equivalence theorem.

Using the same parameter values and design space as in Section 4.1 of \citet{huangejs}, our approach reproduces the reported \(D_s\)-optimal design, which assigns weight \(0.352\) to group size 11 under Reg and weight \(0.648\) to group size 14 under Both. It also yields new designs under other criteria. For example, under \(D\)-optimality, the design assigns Reg at group size 1 with weight \(0.304\) and Both at group size 16 with weight \(0.696\); under \(A\)-optimality, it assigns Reg at group size 1 with weight \(0.169\) and Both at group size 17 with weight \(0.831\). To our knowledge, these designs have not previously been reported.

The equivalence theorem in the Supplementary can be adapted directly to this setting. For the Reg and Gold procedures, one replaces \(\lambda(u_i)\) and \(\bff(u_i)\) by the corresponding expressions in equation (2.4) of \citet{huangejs}. For the Both procedure, the information contribution is obtained by summing \(\lambda_{Bj}(u_i)\) and \(\bff_{Bj}(u_i)\) over \(j=1,2,3\). Thus, the two-assay problem provides a concrete illustration of how the present framework can be extended beyond the single assay setting studied here.   

\subsection{Alternative cost functions}

Throughout this paper, we adopted the linear cost function \(c(x)=1-q+qx,\)
following \citet{huang:2020:cost-GT}. This form captures the basic trade-off between assay cost and subject-enrolment cost and provides a useful starting point for design construction. In practice, however, cost structures may be more complicated. As noted by \citet{turner:2009:cost} and \citet{ma:2023:cost}, pooling and testing costs may exhibit nonlinear behaviour due to batching requirements, administrative overhead, staffing constraints, or other operational considerations.

The proposed methodology is not tied to this linear specification. If a more realistic nonlinear cost function is available, it can be incorporated directly by modifying the cost term in the information matrix. The general design principles remain unchanged, with the main adjustment occurring in the construction of the information matrix itself.

\subsection{Diseases with arbitrary prevalence, sensitivity, and specificity}

Although our motivating application concerns Chlamydia screening, the design framework is not restricted to rare diseases. In principle, it applies to any setting in which the prevalence \(p_0\) lies in \((0,1)\) and the diagnostic assay has nominal sensitivity and specificity values. Group testing is most attractive in low-prevalence settings because the potential savings are greatest there, but the methodology itself does not depend on rarity.

The framework is also flexible in terms of the inferential target. In the \(c\)-optimality setting, the vector \(\bc\) can be chosen to emphasize any linear combination of prevalence, sensitivity, and specificity. In the Chlamydia example, we used \(\bc=(0,1,1)\) to emphasize sensitivity and specificity, but alternatives such as \(\bc=(1,0.5,0.5)\) or \(\bc=(0.5,1,0.5)\) may be more appropriate in other applications. Likewise, \(D_s\)-optimality corresponds to the special case \(\bc=(1,0,0)\), which targets prevalence alone. This flexibility allows the design to be tailored to a broad range of scientific goals, including settings motivated by diagnostic classification, ROC analysis, or cutoff selection.

{\color{black}
\subsection{Sequential and multi-stage group testing}

The proposed framework is naturally connected to sequential and multi-stage group testing procedures, in which an initial estimation stage is followed by downstream operational testing decisions. In this context, the design problem studied here provides the inferential foundation for later stages by yielding accurate estimates of prevalence and assay characteristics before sequential testing is implemented.

The results in Section~\ref{sec:stage2-downstream} illustrate why this initial stage matters in practice. When the parameter values carried forward into Stage II are sufficiently inaccurate, the selected group size may no longer be optimal under the true model, leading to additional assays and positive regret. From this perspective, the contribution of the proposed methodology is not to replace downstream group testing procedures, but rather to strengthen the statistical basis on which they rely.

This perspective is especially relevant in multi-stage screening programmes, where later-stage efficiency depends directly on the quality of the parameter estimates obtained at the outset. A well-designed first stage can therefore improve not only estimation accuracy, but also the operational performance of subsequent testing procedures.}

\subsection{Other design criteria}

Although our numerical illustrations focused on \(D\)-, \(A\)-, \(D_s\)-, \(c\)-, and \(E\)-optimality, the framework is not restricted to these choices. In principle, the same strategy can be adapted to other criteria, provided that the corresponding objective can be handled within the convex optimization framework or through an appropriate reformulation. Examples include criteria such as \(A_s\)-, \(I\)-, and \(L\)-optimality, which may be more appropriate in applications with different inferential priorities.

Accordingly, the methodology developed here should be viewed not as a collection of solutions for a few isolated criteria, but as a general framework for constructing practically relevant group testing designs under a broad range of inferential goals.

\section{Concluding remarks}\label{sec:conclusion}

This paper studies optimal design problems in group testing with a primary focus on precise parameter estimation under imperfect assays and realistic cost constraints. We developed a unified framework for constructing both single- and multi-objective optimal designs, supported by equivalence theorems for approximate designs and modified rounding algorithms for highly efficient exact designs.

The Chlamydia application illustrates the practical value of the framework in a setting where prevalence, sensitivity, and specificity must all be taken into account, and where different inferential priorities can lead to markedly different designs. More broadly, the methodology applies to group testing problems in which accurate parameter estimation is essential for efficient downstream decision-making.

A central message of the paper is that optimal design should be viewed as the first stage of a broader testing pipeline. The designs developed here provide \textbf{Stage I}, in which the underlying model parameters are estimated efficiently, while downstream procedures form \textbf{Stage II}. The numerical results show that this distinction is not merely conceptual: when the parameter values carried forward into Stage II are sufficiently inaccurate, the selected group size may no longer be optimal under the true model, leading to unnecessary additional assays.

The proposed framework accommodates classical single-objective criteria such as \(D\)-, \(D_s\)-, \(A\)-, \(c\)- and \(E\)-optimality, as well as multi-objective criteria that balance competing inferential priorities. It also allows realistic operational constraints to be incorporated directly into the design problem, making the methodology adaptable to a broad class of applied screening settings.  For example, when there are multiple objectives in the group screening study and the objectives have different interests, one can also directly apply the method described in \cite{wong:2023:CVX} to ensure the design delivers user-specified efficiencies with higher efficiencies for the more important objectives.

Several directions are open for future work, including group testing problems with multiple assays, more complex cost structures, and tighter integration with fully sequential or adaptive testing procedures. More broadly, the contribution of this paper lies in providing a principled and practically implementable framework for designing modern group testing studies in which inferential precision and downstream operational efficiency must be addressed jointly.

\bibliographystyle{abbrvnat}
\bibliography{GroupTesting_arXiv_v2}

\end{document}